\documentclass[acmsmall, nonacm]{acmart}

\usepackage{tikz}
\usetikzlibrary{chains,shadows,fit}

\newcommand{\appdirections}{\ref{app:directions}}
\newcommand{\appmaxdist}{\ref{app:maxdist}}
\newcommand{\applightness}{\ref{app:lightness}}
\newcommand{\appcolornames}{\ref{app:color-names}}

\begin{document}

\title{Accessible Color Sequences for Data Visualization}
\author{Matthew~A. Petroff}
\orcid{0000-0002-4436-4215}
\affiliation{%
  \institution{Johns Hopkins University}
  \department{Department of Physics \& Astronomy}
  \streetaddress{3400 N Charles St}
  \city{Baltimore}
  \state{Maryland}
  \postcode{21218}
  \country{USA}
}
\affiliation{%
  \institution{Harvard \& Smithsonian}
  \department{Center for Astrophysics}
  \streetaddress{60 Garden St}
  \city{Cambridge}
  \state{Massachusetts}
  \postcode{02138}
  \country{USA}
}
\email{mpetroff@cfa.harvard.edu}

\begin{abstract}
Color sequences, ordered sets of colors for data visualization, that balance aesthetics with accessibility considerations are presented. In order to model aesthetic preference, data were collected with an online survey, and the results were used to train a machine-learning model. To ensure accessibility, this model was combined with minimum-perceptual-distance constraints, including for simulated color-vision deficiencies, as well as with minimum-lightness-distance constraints for grayscale printing, maximum-lightness constraints for maintaining contrast with a white background, and scores from a color-saliency model for ease of use of the colors in verbal and written descriptions. Optimal color sequences containing six, eight, and ten colors were generated using the data-driven aesthetic-preference model and accessibility constraints. Due to the balance of aesthetics and accessibility considerations, the resulting color sequences can serve as reasonable defaults in data-plotting codes, e.g., for use in scatter plots and line plots.
\end{abstract}

\maketitle



\section{Introduction}

With the rise of electronic publishing starting in the 1990s, scientific visualization, in particular data plotting in scientific publications, switched from primarily using only black and white to primarily using color to represent variables and data. Although these new colorful plots are arguably more aesthetically pleasing, additional problems were introduced due to the somewhat-arbitrary choice of colors. In the case of continuous variables, these shortcomings have been addressed with the introduction of perceptually-uniform colormaps such as \texttt{viridis} \citep{viridis} and \texttt{parula} \citep{parula}, which have begun to replace the use of severely-flawed rainbow colormaps such as \texttt{jet} \citep{jet},\footnote{Note that even improved rainbow colormaps such as \texttt{turbo} \citep{turbo} are not accessible for individuals with color-vision deficiencies.} a subject on which much has been written \citep{Rogowitz1998, Light2004, Crameri2020}. The \texttt{cividis} colormap goes one step further, explicitly optimizing for both individuals with normal color vision and those with color-vision deficiencies \citep{Nuez2018}.

Although there are no reliable global estimates, the prevalence of color-vision deficiencies is as high as 8\% in certain sub-populations \citep{Birch2012}. Human color vision works through the use of long (L), medium (M), and short (S) wavelength color receptors, known as cones, which are sensitive to spectra perceived as red, green, and blue, respectively. Color-vision deficiencies, also commonly known as colorblindness, arise when one---or more---of these cones is either missing or has an altered spectral response. The most common types of color-vision deficiencies are deuteranomaly and protanomaly, which result from altered M and L cones, respectively. The complete lack of M or L cones, respectively known as deuteranopia and protanopia, is also somewhat common. Collectively, these four deficiencies are sometimes referred to as ``red--green colorblindness,'' since they affect the cones responsible for perceiving red or green, although red and green are not commonly confused with each other \citep{Lillo2013}; instead, any color that contains the affected primary color is affected, e.g., protanopes have issues differentiating between blue and purple, since the red that is added to blue to form purple is not perceived. These common color-vision deficiencies are primarily hereditary. As they are caused by recessive genes linked to the X chromosome, they primarily affect men \citep{Neitz2011}. Altered S cone response and the lack of S cones are known as tritanomaly and tritanopia, respectively, but these conditions are much less common, as is monochromacy, the complete lack of color vision \citep{Sharpe1999}.

While colormaps for plotting continuous variables have been improved in recent years based on research into perceptual uniformity, color sequences---ordered sets of colors used for categorical plotting,\footnote{The phrase \emph{color cycles} is also used to refer to such sequences, since plotting codes will cyclically reuse colors in the sequence when the number of items to be plotted exceeds the number of colors in the sequence.} e.g., line plots and scatter plots---have not received similar attention. These colors---or \emph{tones}, as defined by \citet{Bertin1983}---are utilized as a retinal variable to plot distinct components in the same visualization. Particular color sequences have previously been proposed \citep{Brewer2003, Okabe2008}, but they have been created based on a designer's preferences instead of using data-driven derivations. In this paper, the term \emph{color set} is used to refer to a set of colors when their order does not matter, while the term \emph{color sequence} is used when the order does matter. Despite the prevalence of color-vision deficiencies, the default color sequences used in many plotting codes are not accessible to individuals without normal color vision. As a society, we have recognized the need to include individuals with atypical needs through reasonable accessibility accommodations, e.g., curb cuts and closed captioning, and that such accommodations often benefit individuals beyond those that the accommodations were intended to help \citep{Blackwell2016}; there is a need for data-visualization practitioners to make similar accommodations for common color-vision deficiencies.\footnote{Alternative data presentation methods for use by individuals for whom careful color choice is not enough of an accommodation are also important, but they are beyond the scope of the present work.}

In designing an accessible (unordered) color set to be used for general-purpose categorical plotting, cf. for continuous colormaps \citet{Bujack2018}, the following aspects should be considered:
\begin{itemize}
\item A minimum perceptual distance between set colors, including for individuals with color-vision deficiencies, should be enforced to ensure accessibility.
\item A minimum lightness distance between set colors should be enforced to aid with legibility in grayscale printing or display and to aid with accessibility.
\item Very light and very dark colors should be avoided for accessibility and visual weight reasons.
\item Colors that are aesthetically pleasing to a broad audience should be chosen.
\item Colors that can be unambiguously named should be favored, to facilitate verbal and written descriptions.
\end{itemize}
When these colors are used in an ordered color sequence, additional aspects should be taken into account:
\begin{itemize}
\item An ordering that is aesthetically pleasing to a broad audience should be used.
\item Since colors at the end of the sequence will not be used when the number of categories to plot is smaller than the number of colors in the sequence, the colors toward the beginning of the sequence should have larger perceptual and lightness distances between them, to aid with accessibility, and should avoid reusing the same basic color term where possible, to help avoid ambiguity in verbal and written descriptions.
\end{itemize}
As some of these considerations conflict with each other, a balance between them needs to be found, but many of the individual considerations can and will be approached separately. The final result will not be ideal from an aesthetics perspective nor from an accessibility perspective; it aims to be a compromise that accommodates individuals with common color-vision deficiencies while still being palatable to those with typical color vision and being reasonably aesthetically pleasing in a general sense. By accommodating accessibility aspects, the result will also benefit those with normal color vision when plots are viewed in less-than-ideal display conditions, e.g., on washed-out projection screens.

The accessibility concerns can be addressed through technical means by enforcing a minimum distance between colors in a perceptually-uniform color space and by using color-vision-deficiency simulations to also enforce the minimum distance for various color-vision deficiencies. However, a color sequence should also be aesthetically pleasing, to make it easier on the eyes and to encourage adoption of it, which is a matter of personal preference that cannot be directly accounted for with color-perception calculations. To address this, random accessible color sequences were generated and then presented as part of an online survey in an attempt to crowdsource data on aesthetic preference. The results of this survey were then used to train a machine-learning model to quantify aesthetic preference, which was in turn used to rank the randomly-generated color sequences in concert with accessibility constraints in order to select the ones that were both aesthetically pleasing and accessible, including to individuals with color-vision deficiencies.

The remainder of this paper is organized as follows. We start by discussing related work in Section~\ref{sec:relatedwork}, before moving on to cover a procedure for generating random accessible color sets in Section~\ref{sec:colorsets}. Next, we discuss a color-sequence survey that was used to collect aesthetic-preference data in Section~\ref{sec:survey}, a model constructed using the data in Section~\ref{sec:model}, and results from the model and other color-sequence-accessibility constraints in Section~\ref{sec:results}. Finally, we provide a discussion of the results in Section~\ref{sec:discussion}, before making a comparison to existing color sequences in Section~\ref{sec:comparison} and concluding in Section~\ref{sec:conclusions}.

\section{Related Work}
\label{sec:relatedwork}

Previous work on color sets for visualization can primarily be categorized as designer color sequences, tools for creating custom color sets, and procedures for producing fixed color sequences that are optimal by a particular metric. There is also literature on adapting colors to a particular visualization, such as by using semantically-meaningful colors \citep{Setlur2016} or by accounting for feature adjacency or overlap \citep{Lu2021, Yuan2022}, but such work will not be considered here further, since the goal of the present work is to devise generally-applicable color sequences, although both \citet{Lu2021} and \citet{Yuan2022} are notable for applying machine learning techniques. The most well-known work in the designer category is the \emph{ColorBrewer} tool \citep{Brewer2003}, which presents a series of hand-crafted color sequences of various types, which were designed by an expert, primarily for use with visualizing data on a map. Also in this category are \citet{Okabe2008} and \citet{Tol2021}, which present color sequences that were hand-crafted to maintain accessibility for color-vision-deficient individuals, and \citet{Zeileis2009}, which presents a procedure for constructing color sequences.

Two notable examples in the category of tools for creating custom color sets are the \emph{I Want Hue} tool \citep{iwanthue}, which uses a rule-based approach for guiding custom color set creation, and the \emph{Colorgorical} tool \citep{Gramazio2017}, which combines a set of adjustable parameters with some degree of stochasticity to generate custom color sets that attempt to be both aesthetically-pleasing and accessible. Earlier rule-based tools include those described in \citet{Bergman1995}, \citet{Healey1996}, \citet{Meier2004} and \citet{Hu2012}, the latter two of which are not specifically targeted at data visualization. Other tools not specifically targeted at data visualization are those of \citet{ODonovan2011} and \citet{Mellado2017}. In \citet{ODonovan2011}, the authors developed an aesthetic-preference model based on color themes published in two different online communities for graphic designers to share such themes as well as survey data; they then used this model to develop tools for optimizing an input color sequence, for extracting a color sequence from an image, and for suggesting additional colors to extend a partial color sequence. In \citet{Mellado2017}, a framework for interactive optimization of existing palettes is presented.

Finally, a significant example of optimizing to a metric is \citet{Glasbey2007}, which alternatively used sequential search or simulated annealing to create color sequences that maximized the distance between colors in the CIELAB color space. Previously, \citet{Campadelli1999} had used a graph-based algorithm to find a color set that maximized the distance between colors in the CIELUV color space but with a more limited set of input colors. Earlier work with a similar goal---but published prior to the development of perceptually-uniform color spaces---is that of \citet{Kelly1965}. Both \citet{Troiano2008} and \citet{Yanagida2014} presented algorithms for modifying existing color sequences to improve their accessibility to color-vision-deficient individuals.

Unfortunately, much of this previous work has generally not considered accessibility for color-vision-deficient individuals, although some work beyond visualization, such as choosing palettes for web design \citep{Tigwell2017}, has done so. Additionally, the existing data-driven approaches have generally not attempted to simultaneously account for both aesthetic preference and accessibility.

\section{Generating Color Sets}
\label{sec:colorsets}

In order to create accessible color sets, a minimum perceptual distance between colors must be enforced. The perceptual distance between two colors is calculated by determining the Euclidean distance between the two colors, $\Delta E'$, in the CAM02-UCS perceptually-uniform color space \citep{Fairchild2013, Luo2013}. This color space was designed such that a Euclidean distance of one between two color coordinates corresponds to an equal perceptual color difference throughout the color space for individuals with normal color vision. Note that observing conditions affect color perception, so the color-difference distance that corresponds to a just-noticeable difference depends on the observing conditions, including the size of the colored regions \citep{Szafir2018}. Regardless of the just-noticeable-difference color distance, larger distances correspond to easier-to-differentiate colors, up to a point \citep{Carter2010}, and these easier-to-differentiate colors are more accessible.

In order to account for color-vision deficiencies, the method presented by \citet{Machado2009} is used to simulate the appearance of colors for color-vision-deficient individuals,\footnote{The method presented in \citet{Machado2009} does not specify how its simulation matrices should be applied to RGB colors. This was noted by \citet{Harding2020}, which also notes that the figures in \citet{Machado2009} were created by applying the simulation matrices to gamma-encoded sRGB values. Here, we apply the matrices to linear sRGB values, as was also done by \citet{Nuez2018} and \citet{Harding2020}.} and CAM02-UCS is then used to calculate and enforce a minimum perceptual distance for the transformed colors, with separate simulations performed for protanopia, deuteranopia, and tritanopia. The method presented in \citet{Machado2009} allows for simulating both dichromacy, the complete lack of one cone, and various severities of anomalous trichromacy, the shift in spectral response of a cone. The method was chosen due to this support of anomalous trichromacy, as well as since the method is physiologically-motivated and since the authors validated it to some degree with experimental data. Additionally, the same simulation and perceptual-distance calculation techniques were used in \citet{Nuez2018} for creating the \texttt{cividis} colormap, except \citet{Nuez2018} did not consider tritanopia. More specifically, we define the distance metric $\Delta E_\text{cvd}$ for colors $c_1$ and $c_2$ as
\begin{equation}
\label{eq:cvd}
\Delta E_\text{cvd}(c_1, c_2) = \min_{s\in\text{sev.}}\min\begin{Bmatrix}
\Delta E'[c_1,c_2] \\
\Delta E'[\text{deut}_s(c_1), \text{deut}_s(c_2)] \\
\Delta E'[\text{prot}_s(c_1), \text{prot}_s(c_2)] \\
\Delta E'[\text{trit}_s(c_1), \text{trit}_s(c_2)]
\end{Bmatrix},
\end{equation}
where $\Delta E'$ is the CAM02-UCS color distance and $\text{deut}_s$, $\text{prot}_s$, and $\text{trit}_s$ are the color-vision-deficiency simulation results of \citet{Machado2009} for severity $s$ for deuteranomaly, protanomaly, and tritanomaly, respectively. Unless otherwise noted, all integer severities are included, $\{1\ldots100\}$.
In addition, a minimum lightness distance, $\Delta J'$, is enforced to maintain accessibility for individuals with monochromacy and for accommodating grayscale printing. While calculations are performed using CAM02-UCS, colors are defined using 8-bit integer sRGB colors.

\subsection{Additional accessibility constraints}
\label{sec:additionalconstraints}

Additional accessibility aspects also need to be considered, specifically maximum lightness and compatibility with grayscale printing. A maximum lightness value needs to be set to maintain sufficient contrast with a white background. While no minimum-contrast standards exist for visualizations, such standards do exist for text, specifically the W3C Web Content Accessibility Guidelines 2.1 standard \citep{WCAG}, but these guidelines are not a good fit for visualizations.\footnote{The guidelines also define color contrast in terms of the (linear) sRGB color space, which is not perceptually uniform.} As only two colors need to be considered for text content---the text color and the background color---much-stricter contrast guidelines can be used. Thus, following these guidelines, even those meant for large heading text, eliminates the use of most lighter colors. With fewer possibilities, the resulting color set must have a smaller minimum perceptual distance, which is problematic.

Fortunately, experimental data on how lightness affects plot readability has been collected, specifically by \citet{Smart2019}, whose authors conducted an experiment where two scatter-plot markers of the same color and size but potentially of different shapes were presented along with gray distractor markers to research-survey respondents. The respondents were asked to answer as quickly and accurately as possible, and response times and accuracy were recorded, for markers of various CIELAB lightness, $L^*$.\footnote{CIELAB lightness, $L^*$, is approximately equivalent to CAM02-UCS lightness, $J'$.} In their analysis, the authors considered how lightness affects accuracy, finding that it decreased for $L^*>92$. While only the use of extremely-light colors affected accuracy, darker---but still light---colors may also make it difficult to discern marker shape and thus negatively affect the readability of a scatter plot; very dark and very light colors are generally poor choices for data visualization \citep{Tufte1990}. The authors of \citet{Smart2019} published their raw data, allowing for a reanalysis of it in terms of how lightness affects response time; this reanalysis is described in Appendix~\applightness. It found that for 0.25\textdegree-size markers, response time increased for $L^*>84.6$. As this value should be used as a minimum requirement and since color sets with fewer colors allow for more flexibility, stronger restrictions with increased readability can be placed on color sets with fewer colors.

Although the primary use-case for the color sequences being discussed here is use with emissive color-capable digital display devices, grayscale is a valid alternative display mode that still needs to be considered. Grayscale printing or display is handled by the lightness-difference restriction for monochromacy. Color printing was not specifically considered, since the printable gamut depends on the printing process and paper and thus results in much variability in appearance; however, the grayscale considerations also apply to color printing, so this omission is not a significant issue.

A final consideration for accessibility is the interaction between color perception and language, but unlike the previously-discussed color- and lightness-distance considerations, the interaction with language is a softer constraint, which mostly affects communication about color. Thus, it will only be used as part of the ranking criteria used to generate the final results, so discussion of this is aspect is postponed to the discussion of the final results in Sections~\ref{sec:final-sets} and \ref{sec:final-sequences}.

\subsection{Random Set Creation}
\label{sec:randomset}

In order to randomly generate accessible color sets, a rejection-sampling-based method is used. This process is agnostic to the exact minimum perceptual distance, and different constraints will be used when differently-sized color sets are desired, since a larger minimum perceptual distance can be enforced when fewer colors are needed. Since sRGB to CAM02-UCS conversions are computationally expensive, but the $\sim$16.8 million possible 8-bit sRGB colors easily fit in memory, the CAM02-UCS colors are precomputed for every possible color, both for normal color vision and for the three types of dichromacy.

To generate a color set, a starting color is chosen at random. Then, each possible color is checked to see if it is far enough away in both lightness and perceptual distance, both for normal color vision and for color-vision deficiencies, at the maximum color-vision deficiency severity, dichromacy, i.e., $\Delta E_\text{cvd}$ is calculated for only a single severity, 100. Of these remaining colors, one is chosen at random using rejection sampling in the CAM02-UCS color space. The process is then repeated until the color set contains the desired number of colors or no colors matching the distance criteria remain in the sRGB gamut. Checking each possible color prior to rejection sampling is done as this method is guaranteed to complete in finite time---unlike rejection sampling alone---since rejection sampling is not performed when there are no remaining candidate colors. After the color set is generated, it is checked at intermediate levels of color-vision-deficiency severity to ensure that the minimum-perceptual-distance requirement is met there as well, i.e., $\Delta E_\text{cvd}$ is calculated with severities $\{1\ldots100\}$. If the distance requirement is not met, the color set is thrown out. Checking a coarse color-vision-deficiency interval during set generation was tried but removed, since the performance penalty outweighed the gains from having to try again fewer times.

With this method in place, it is now possible to randomly generate color sets of various sizes that meet various minimum-perceptual-distance and minimum-lightness-distance requirements. Color sets containing six, eight, and ten colors were generated using this method, with minimum perceptual distances of 20, 18, and 16, respectively. The minimum perceptual distance constraints were chosen as a compromise between maximizing accessibility and allowing for an expanded range of different color sets in order to allow for more leeway for aesthetic preference. The choice of the constraints was informed by an analysis that created a maximally-distant color sequence via a sequential-search-algorithm method, which is described in Appendix~\appmaxdist.

Two groups of color sets were generated, with 10k of each size for each group. The first used a minimum lightness distance of 2 for all three set sizes and a lightness range of $J' \in [40, 90]$, where $J'$ is the CAM02-UCS lightness value, truncating the gamut to $\sim$13.1 million colors. These values were chosen to eliminate sets with egregiously-bad accessibility but not excessively constrain the sets. The maximum minimum-perceptual distances among the generated sets of color sets were 24.3, 20.2, and 18.0 for the configurations with six, eight, and ten colors, respectively.

For the second group of color sets, tighter constraints were used. Based on the previously-mentioned lightness analysis, which favored $L^*<84.6$,  the minimum color distances for each set length are also used for minimum lightness distance from the white background for this group of color sets, with lightness ranges of $J'\in[40, 80]$, $J'\in[40, 82]$, and $J'\in[40, 84]$ used for color sets of length six, eight, and ten, respectively. For improved accessibility in grayscale, the minimum lightness difference was increased to allow a maximum of three additional colors in the lightness range, resulting in a minimum $\Delta J'$ of 5.0, 4.2, and 3.6 for sets of six, eight, and ten colors, respectively. With these additional restrictions, an additional group of color sets were generated. The maximum minimum-perceptual distances among these new sets of color sets were 23.6, 19.6, and 16.9 for the configurations with six, eight, and ten colors, respectively.

The first group of color sets, with the looser constraints, were used for the aesthetic preference survey described in the next section, while the second group of colors sets, with the tighter constraints, were used for the final results, described in Section~\ref{sec:final-sets}.\footnote{Due to a bug in the original version of the color-set-generation code that caused out-of-gamut colors to be wrapped into the sRGB gamut, the rejection sampling done for the first group of sets was not done strictly in CAM02-UCS. This bug was identified and fixed before the second group of sets were generated.} As the second group of color sets were drawn from a gamut that is a subset of the original gamut, any modeling of the aesthetics of the first group can be used without extrapolation on the second group. In retrospect, it likely would have been better to have used the second, tighter set of constraints for the aesthetic preference survey as well, but the author had a naive hope that the survey would receive significantly more responses than it did, such that the looser constraints would have made the results more-generally useful.

\section{Color-sequence Survey}
\label{sec:survey}

While one could take an ontological approach to modeling aesthetic preference by trying to define a set of rules that make a pleasing color sequence, as is done by \emph{I Want Hue} \citep{iwanthue}, such a method is error-prone and substantially biased toward the personal preferences of the drafter of the rules. Instead of using ontologies, an alternative approach that is gaining traction in many fields is to infer a pattern from a large dataset using machine-learning techniques. This is the approach pursued here.

\begin{figure*}
\centering
\frame{\includegraphics[width=0.95\textwidth]{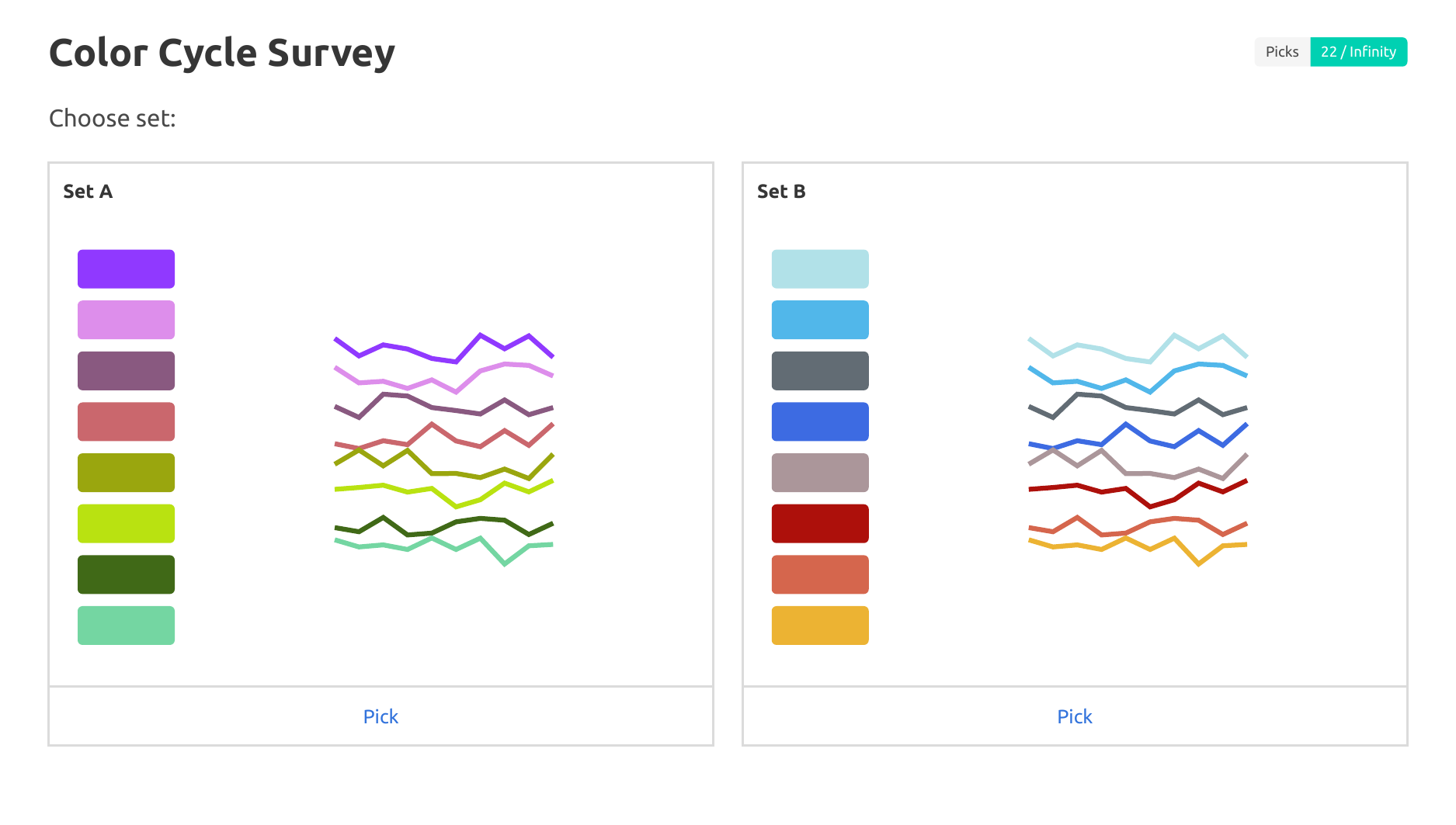}}
\caption{Color survey sets interface. The survey respondent was presented with two separate color sets and asked to choose the more pleasing set; after this step, four possible orderings were presented, and the respondent was asked to choose the most pleasing ordering. The plot rendering was randomly changed between a line plot and a scatter plot, and different line thicknesses and marker sizes were used. The rendering was not shown on smaller screens and was not shown when picking orderings.}
\label{fig:surveysets}
\Description[Screenshot of color survey sets interface]{A screenshot showing the color survey sets interface is presented. It shows two boxes, each with a selection button. Each box contains eight colored boxes and a line plot with eight lines of the same colors; these colors represent the colors in a specific color set. In the top right corner, there is a running count of the number of responses the user has provided.}
\end{figure*}

To this end, an online color-sequence survey was created.\footnote{While active, the survey was accessible at: \url{https://colorcyclesurvey.mpetroff.net/}} After presenting the user with an introduction, a color-vision-deficiency questionnaire, and directions and after obtaining consent for the data collection, the primary survey started. The introduction and directions are reproduced in Appendix~\appdirections. In it, the user was presented with two of the color sets described in the previous section and was asked to choose the one that was, in the user's opinion, more aesthetically pleasing as a gestalt. Then, four random orderings of the chosen set were displayed, and the user again made a selection to taste. This basic process was then repeated again and again, with sets of either six or eight colors; there was no fixed end point. When the survey was first deployed, sets with ten colors were also presented, although this option was removed after a short period of time to increase the number of responses for the other two set sizes. For the choice of color set, each set was presented ordered by hue angle, since this makes the two sets easier to compare than if they were randomly ordered (or ordered by RGB values). Only two sets were presented to make for an easier choice. Additionally, a line-plot or scatter-plot rendering was shown with each color set when screen space was available, i.e., not on mobile devices. The line thicknesses and marker sizes were randomly varied. A screenshot of the survey page is shown in \figurename~\ref{fig:surveysets}.

For the choice of ordering, four orders were presented, since multiple orderings are easier to compare than multiple sets, since the colors are all the same. Asking the user to order the colors to taste, instead of presenting possible orderings, was considered, but it was decided that while such an approach yields more information per response, it takes much longer and requires more effort, so each user would likely respond many fewer times. Thus, the simpler approach of presenting four possible ordering was taken.

The survey operated for approximately two years, from December 2018 through December 2020. It was promoted in user communities, primarily that of Matplotlib / Python, and was also linked to from banners on the author's personal website\footnote{\url{https://mpetroff.net/}} and that of the Pannellum panorama viewer \citep{Petroff2019}. Participation was entirely voluntary, and no compensation was offered. The survey collected $\sim$22k responses from $\sim$2.2k user sessions. More specifically, there were 10\,347 six-color, 10\,371 eight-color, and 1705 ten-color responses recorded. For 5.5\% of the user sessions, the respondent self-identified as having some form of color-vision deficiency. Based on partial IP addresses, the countries, in decreasing order, with the largest number of user sessions were the United States, Germany, the United Kingdom, and Japan;\footnote{The 2019-12-24 release of the MaxMind GeoLite2 Country geolocation database was used to map partial IP addresses to countries.} these four countries accounted for just over half of the user sessions.

\section{Modeling Aesthetic Preference}
\label{sec:model}

In order to create a quantified measure of aesthetic preference, models must be constructed such that they can rank sets of color sets or color sequences, and model parameters must be able to be inferred from the pairwise responses collected using the previously-discussed survey. Ideally, these models should take a single color set or color sequence as an input and produce a continuous numerical output score, since this results in a simplified ranking process when compared to a model that takes two inputs and produces a binary-preference output. Additionally, a single model that works for a range of different color set or sequence sizes would be preferred, since it could be generalized to additional sizes. This would make it necessary to have only two models, one for scoring color sets and one for scoring color sequences for a given color set. To this end, a model was designed using an artificial-neural-network technique. Artificial neural networks consist of a set of units connected with trainable weights and nonlinear transformations and, given a sufficient number of parameters, can be used to approximate arbitrary functions \citep{Russell2010}.

To meet the requirement that the model can be trained using pairwise data while still producing a continuous numerical output score, a conjoined network architecture \citep{Bromley1993} was chosen. In such an architecture, the outputs of two identical copies of a network with shared weights are connected together for the training process, in this case with a subtraction operation. Each copy takes one input from the training pair, and the result of the subtraction operation joining the outputs of the two copies is used to train the model using a binary-cross-entropy loss function. This effectively trains each copy to produce a score for its input. Once the training process is complete, the joining operation is removed, and a single copy is used with a single input to produce an output score.

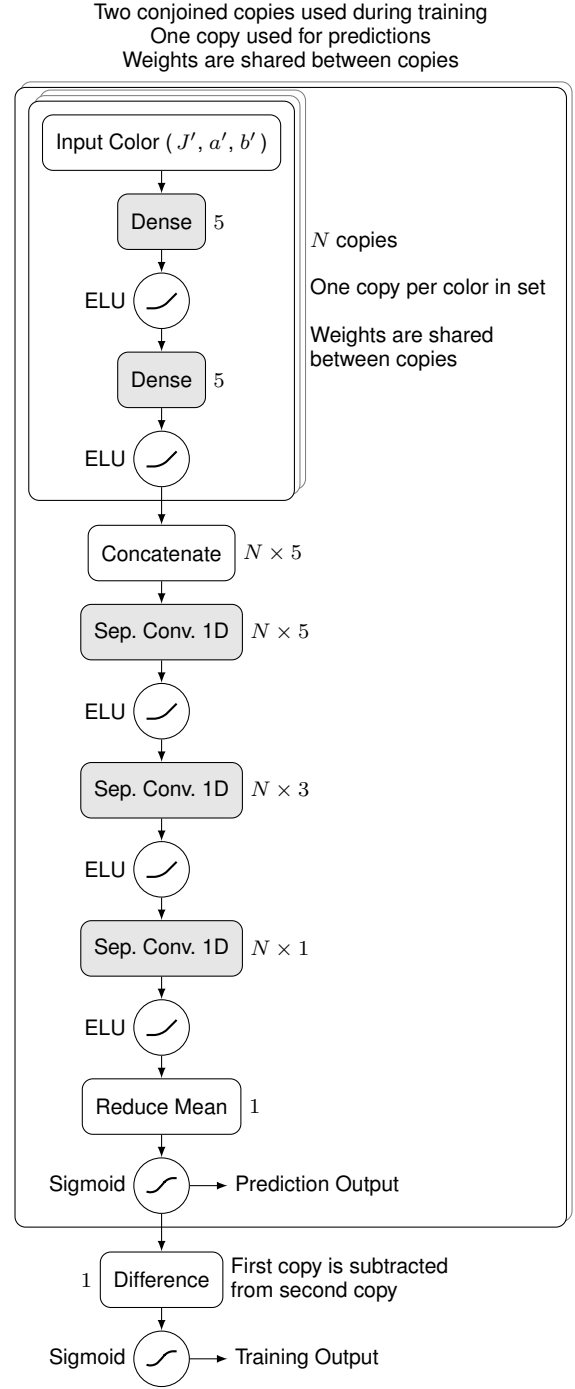
\begin{figure}
\centering
\pgfdeclarelayer{background0}
\pgfdeclarelayer{background1}
\pgfsetlayers{background0,background1,main}
\begin{tikzpicture}
\sffamily
\tikzset{>=latex}
\footnotesize

\tikzset{
multi/.style={fill=white,double copy shadow={draw=gray}},
double/.style={fill=white,copy shadow={draw=gray}},
netnode/.style={draw, minimum size=5ex,inner sep=5pt,rounded corners},
netnodecircle/.style={draw, circle, minimum size=5ex},
learned/.style={fill=gray!20}
}

\tikzset{elu/.pic={
\draw[-,thick,anchor=center,rounded corners=1mm] (-0.2,-0.1)to(0,-0.1)to(0.2,0.1);
}}

\tikzset{sigmoid/.pic={
\draw[-,thick,anchor=center,rounded corners=0.7mm] (-0.2,-0.08)to(-0.1,-0.1)to(0,0)to(0.08,0.1)to(0.2,0.1);
}}

\node[netnode] (incolor) {Input Color (\,$J'$, $a'$, $b'$\,)};
\node[netnode, learned, below = 3mm of incolor] (dense1) {Dense};
\node[netnodecircle, below = 3mm of dense1, path picture={\pic{elu};}] (act1) {};
\node[netnode, learned, below = 3mm of act1] (dense2) {Dense};
\node[netnodecircle, below = 3mm of dense2, path picture={\pic{elu};}] (act2) {};
\node[netnode, below = 5mm of act2] (concat) {Concatenate};
\node[netnode, learned, below = 3mm of concat] (conv1) {Sep. Conv. 1D};
\node[netnodecircle, below = 3mm of conv1, path picture={\pic{elu};}] (act3) {};
\node[netnode, learned, below = 3mm of act3] (conv2) {Sep. Conv. 1D};
\node[netnodecircle, below = 3mm of conv2, path picture={\pic{elu};}] (act4) {};
\node[netnode, learned, below = 3mm of act4] (conv3) {Sep. Conv. 1D};
\node[netnodecircle, below = 3mm of conv3, path picture={\pic{elu};}] (act5) {};
\node[netnode, below = 3mm of act5] (mean) {Reduce Mean};
\node[netnodecircle, below = 3mm of mean, path picture={\pic{sigmoid};}] (act6) {};
\node[netnode, below = 5mm of act6] (diff) {Difference};
\node[netnodecircle, below = 3mm of diff, path picture={\pic{sigmoid};}] (act7) {};

\node[right = 5mm of act6] (predout) {Prediction Output};
\node[right = 5mm of act7] (trainout) {Training Output};

\path[style={draw, ->}] (incolor) edge (dense1);
\path[style={draw, ->}] (dense1) edge (act1);
\path[style={draw, ->}] (act1) edge (dense2);
\path[style={draw, ->}] (dense2) edge (act2);
\path[style={draw, ->}] (act2) edge (concat);
\path[style={draw, ->}] (concat) edge (conv1);
\path[style={draw, ->}] (conv1) edge (act3);
\path[style={draw, ->}] (act3) edge (conv2);
\path[style={draw, ->}] (conv2) edge (act4);
\path[style={draw, ->}] (act4) edge (conv3);
\path[style={draw, ->}] (conv3) edge (act5);
\path[style={draw, ->}] (act5) edge (mean);
\path[style={draw, ->}] (mean) edge (act6);
\path[style={draw, ->}] (act6) edge (diff);
\path[style={draw, ->}] (diff) edge (act7);

\path[style={draw, ->}] (act6) edge (predout);
\path[style={draw, ->}] (act7) edge (trainout);

\node[right = 0mm of dense1] {$5$};
\node[right = 0mm of dense2] {$5$};
\node[right = 0mm of concat] {$N \times 5$};
\node[right = 0mm of conv1] {$N \times 5$};
\node[right = 0mm of conv2] {$N \times 3$};
\node[right = 0mm of conv3] {$N \times 1$};
\node[right = 0mm of mean] {$1$};
\node[align=left, right = 0mm of diff] {First copy is subtracted\\from second copy};
\node[left = 0mm of diff] {$1$};
\node[left = 0mm of act1] {ELU};
\node[left = 0mm of act2] {ELU};
\node[left = 0mm of act3] {ELU};
\node[left = 0mm of act4] {ELU};
\node[left = 0mm of act5] {ELU};
\node[left = 0mm of act6] {Sigmoid};
\node[left = 0mm of act7] {Sigmoid};

\begin{pgfonlayer}{background1}
\node (colorbox) [netnode, multi, fit = (incolor) (act2)] {};
\end{pgfonlayer}
\node[align=left, right = 15mm of act1] (colorboxlabel) {$N$ copies\\~\\One copy per color in set\\~\\Weights are shared\\between copies};

\begin{pgfonlayer}{background0}
\node (conjbox) [netnode, double, fit = (colorbox) (colorboxlabel) (act6)] {};
\end{pgfonlayer}
\node[align=center, above = 1mm of conjbox] (conjlabel) {Two conjoined copies used during training\\One copy used for predictions\\Weights are shared between copies};

\end{tikzpicture}
\caption{Artificial-neural-network architecture overview. Shading denotes nodes with trainable parameters, and the numbers next to nodes denote their output dimensions. The 1D separable convolutions use a kernel of size 5, and the inputs are zero-padded before the kernel is applied such that the output shape is identical to the input shape. For the set model, the input colors were ordered along one of the three CAM02-UCS axes, and a separate copy of the model was independently trained for each ordering; the outputs of the three copies were then averaged. For the sequence model, the input colors were ordered per the sequence ordering.}
\label{fig:ann}
\Description[Graphical representation of artificial-neural-network architecture]{A graphical representation of the artificial-neural-network architecture described in the text is shown. At the top of the figure, a box contains an "input color" node, followed by a "dense" node with the numeral 5 next to it, an "ELU" node, a "dense" node with the numeral 5 next to it, and an "ELU" node. The box is labeled to indicate that there are N copies of it, that there is one copy per color in the color set, and that weights are shared between copies. The box is followed by a "Concatenate" node labeled with "N x 5", a "Sep. Conv. 1D" node labeled "N x 5", an "ELU" node, a "Sep. Conv. 1D" node labeled "N x 3", an "ELU" node, a "Sep. Conv. 1D" node labeled "N x 1", an "ELU" node, a "Reduce Mean" node with the numeral 1 next to it, and a "Sigmoid" node. Next to the "Sigmoid" node is an arrow pointing to a "Prediction Output" label. The previously-mentioned nodes are all contained in a box labeled to indicate that there are two conjoined copies used during training, that there is one copy used for predictions, and that weights are shared between copies. The box is followed by a "Difference" node with the numeral 1 next to it and labeled to indicate that the first copy is subtracted from the second copy. This node is followed by a "Sigmoid" node, which is followed by an arrow pointing to a "Training Output" label.}
\end{figure}

To allow a single model to be used with different input sizes, the model was designed to use a hybrid convolutional architecture. The individual colors in a color set or sequence are passed to the network as CAM02-UCS coordinates, divided by 100 for normalization. Each color is then passed through two dense, fully-connected network layers, each with five nodes. The weights for these layers are shared between each input color, so the number of input colors does not affect the number of weights in this portion of the network. These layers can be thought of as a way of learning an optimal encoding for the input colors. The outputs of the fully-connected layers are then passed through a set of three depthwise-separable convolution layers, with five, three, and one output channels, respectively; a kernel size of 5 is used, and the inputs are zero-padded before the kernel is applied such that the output shape is identical to the input shape. As these layers perform convolutions, they again do not depend on the number of input colors. Both the fully-connected layers and the convolution layers are each followed by an exponential linear unit (ELU) non-linear activation function \citep{Clevert2016}. The output of the final convolution layer is averaged and then passed through a sigmoid activation function, producing an output score. This is repeated for the second half of the conjoined network, and the two scores are differenced and passed through a final sigmoid activation function. A graphical overview of this architecture, which contains 167 trainable parameters, is presented in \figurename~\ref{fig:ann}.

The same model architecture is used for handling both the set data, where ordering does not matter, and the sequence data, where the ordering does matter. However, separate model instances are used. For the sequence data, there is one obvious choice for ordering the input data, the original sequence order, so this is what is used. The choice of ordering for the set data is not as straightforward. In the end, three orderings were used, corresponding to the set colors sorted along the three CAM02-UCS axes. A separate model instance was independently trained using each ordering on the same data, and the resulting output scores are averaged. This method was used as it was found to produce better results than using any of the axes individually or by ordering the colors based on hue angle.\footnote{When the data were ordered by hue angle, the padding for the convolution layers was modified to cyclically wrap to avoid boundary effects.} The three-model ensemble has a total of 501 trainable parameters.

Ideally, the number of pairwise responses would greatly exceed the number of color sets with which to train the model. However, the number of responses is roughly equal to the number of color sets, and the response data have significant variance given that aesthetic preference is subjective and not always clear. Thus, care must be taken to maximize the amount of information extracted from the survey data while preventing overfitting of the model and minimizing the effect of the data split and initialization values on the training process. To this end, bootstrap aggregation \citep{Breiman1996}, i.e., bagging, was used to train a homogeneous ensemble of 100 models in parallel. For each model, training data were randomly drawn from the responses with replacement, creating 100 distinct training sets while still using all of the available data and allowing for overfitting to be checked for; each training set had a number of entries equal to 80\% of the total number of survey responses for sets containing six colors. For each model, a test set was created using the responses not used in the training set. The outputs of the ensemble are averaged to produce the final output score. Data with six colors and with eight colors were used simultaneously during training but were treated separately when creating the training and test splits. Data with ten colors were not used.

The model ensembles were each trained, in a deterministic manner, for 120 epochs using the RMSprop optimization algorithm \citep{Tieleman2012} with a learning rate of 0.01 and a batch size of 1024; L2 activity regularization was applied to all parameters with a value of 0.0001. Two separate model ensembles were trained, one for color sets and one for color sequences. In the case of color sequences, each response was converted from being a pick of the best of four possible orderings into three pairwise responses, i.e., the preferred ordering paired with each of the three worse orderings.

\section{Results}
\label{sec:results}

\subsection{Performance of Model}

On the train set, the final set model achieved accuracies of 58.7\% and 58.0\% for the six-color and eight-color data, respectively. On the test set, it achieved accuracies of 57.8\% and 57.0\% for the six-color and eight-color data, respectively. The overall accuracies on the entire dataset were 58.2\% and 57.5\% for the six-color and eight-color data, respectively.\footnote{This is not quite the same as the weighted average of the train- and test-set accuracies due to the stochastic nature of bagging.}

The final sequence model achieved train-set accuracies of 55.7\% and 55.8\% for the six-color and eight-color data, respectively. On the test set, it achieved accuracies of 54.3\% and 54.7\% respectively. The overall accuracies on the entire dataset were 55.0\% and 55.3\% for the six-color and eight-color data, respectively.

As previously mentioned, sets with ten colors in them were presented as part of the color-sequence survey originally but were removed fairly early into the survey period, resulting in many fewer responses. As the training procedure required equal numbers of sets or sequences of different sizes, the ten-color data were excluded during training. However, this presented an opportunity for validating the model. After the model was finalized and its design, hyperparameter selection, and training were finished, the set and sequence models were evaluated on the ten-color data. With these data, the set model achieved an accuracy of 57.5\%, identical to that achieved for the eight-color data. The sequence model achieved an accuracy of 54.7\%.

\subsection{Validation of Model}

As a cross-check on the machine-learning-derived color-set model, it was compared to a linear model based on summary statistics. The minimum, mean, and maximum of the chroma\footnote{Chroma is a measure of colorfulness and is related to---but not identical to---saturation.} and lightness values of each color set were used. Best-fit model parameters were inferred using a least-squares regression on the entirety of the survey data, where the differences between the summary statistics for the two sets in each response pair were regressed against the binary preference value. This simple model had accuracies of 53.9\%, 54.8\%, and 54.6\% for the six-color, eight-color, and ten-color sets, respectively, which are significantly\footnote{$p \ll 0.01$, $p \ll 0.01$, and $p \approx 0.046$ for the six-color, eight-color, and ten-color sets, respectively, based on a one-sided, two-sample $Z$ test \citep{Zou2003}.} less than the accuracies of the machine-learning model. Based on the linear model, the three most-significant summary statistics were, starting with the most significant, mean chroma, maximum chroma, and maximum lightness; higher mean chroma, lower maximum chroma, and lower maximum lightness values were the preferred trends for these summary statistics.

An additional comparison was made with the pair-preference model used as a proxy for color-set aesthetic preference in \citet{Gramazio2017} and based on data collected by \citet{Schloss2010}. This model scores color pairs using a linear-regression model based on coolness, hue similarity, and lightness contrast. To extend the pair-preference scores to a full color set, pair scores are calculated for each possible color pair in the set. In \citet{Gramazio2017}, the color-set score was determined by using the minimum preference score of all the pairs. When applied to the color-set survey data collected in the present study, this method was found to offer little to no predictive power, with accuracies of 49.8\%, 49.2\%, and 50.1\% for the six-color, eight-color, and ten-color sets, respectively. Calculating and using the mean of the pair-preference scores instead of the minimum, however, resulted in accuracies of 52.4\%, 52.7\%, and 54.4\% for the six-color, eight-color, and ten-color sets, respectively.

Finally, a verification survey was also used. The original color-sequence survey was modified to operate in a fully-deterministic manner, such that the same color sets were shown in the same order every time and such that the color sets used for the ordering questions were also deterministic instead of being the user's chosen color sets. Each user was asked to answer 100 question pairs, where the second 50 pairs were identical to the first 50 pairs. By asking the same questions to each user, the consistency between different individuals can be evaluated, and by asking each user the same questions twice, the internal consistency of individuals can be evaluated. Seven users completed this verification survey, for a total of 700 response pairs.

For color sets, the average internal consistency for users was 63\%, and the average consistency between a given user and the mode of the other users' responses was 58\%.\footnote{The mode was calculated separately for each individual user with that user's responses excluded to avoid bias. A given response was considered half correct if the mode was a tie.} On these data, the machine-learning model correctly predicted the individual responses 59\% of the time and correctly predicted the most common response 70\% of the time. For sequences, the average internal consistency for users was 43\%, and the average consistency between a given user and the mode of the other users' responses was 31\%.\footnote{If the mode was a tie and the response matched one of the tied answers, the response was considered fractionally correct, based on the number of tied answers.} Note that there were four choices for the sequence questions, so the expected consistency for random responses would be 25\% instead of 50\%. On these data, the machine-learning model correctly predicted the individual responses 32\% of the time and correctly predicted the most common response 37\% of the time. With the caveat of the limited sample size, these data show that the machine-learning model performs similarly to a single individual in its ability to predict the responses of other individuals, and its accuracy improves when predicting the most common response, as the response data are noisy. Given that the sequence data have a larger discrepancy between the internal consistency for users and the inter-user consistency than the set data, this suggests that the sequence data are inherently noisier, which could explain the lower accuracy of the machine-learning model on the sequence data.

\subsection{Final Color Sets}
\label{sec:final-sets}

As mentioned in Section \ref{sec:randomset}, a second group of color sets with tighter accessibility constraints were also generated, and the remainder of this paper will focus on those sets. These sets were each scored using the trained neural-network color-set model, resulting in score ranges of [0.29, 0.96], [0.28, 0.95], [0.29, 0.93], for six, eight, and ten colors, respectively.

One last consideration to take into account is the interaction between color perception and language, since the easier a color is to name, the easier it is to refer to the color in a verbal or written description \citep{Stone2003}. The authors of \citet{Heer2012} developed a probabilistic naming model for English based on several million crowdsourced color value--name pairs from the xkcd color survey \citep{Munroe2010}. In addition to naming given color values, this model can also produce color-saliency scores, which quantify the degree to which a color value can be uniquely named. A prototypical example of a commonly-used color name should have a high color-saliency score, while a color value that is on the transition between two commonly-used color names, e.g., halfway between pink and purple, should have a low color-saliency score. However, the existing model uses 153 color names, far more than the $\sim$30 distinct names commonly used by untrained subjects \citep{Derefeldt1995} or the 27 categorically-distinct regions of categorically-similar colors shown to be contained in the sRGB gamut \citep{Griffin2019}, and no attempt to merge color synonyms was made in its construction. As the existence of rarely-used color names or alternative color names does not affect the nameability of a color value in the same way that being on the transition between two commonly-used color names does, the large number of color names and lack of synonym merging negatively affects that usefulness of the existing model's color-saliency metric. To remedy these shortcomings, the model was modified to merge color synonyms and remove rarely-used color names based on the Hellinger distance \citep{LeCam2000} between color-name probability distributions, a process that is described in Appendix~\appcolornames. The revised model was then used to score each of the previously-generated color sets, resulting in score ranges of [0.29, 0.76], [0.32, 0.71], [0.34, 0.69], for six, eight, and ten colors, respectively.

The color-saliency scores were found not to correlate with the neural-network color-set scores. Final color-set scores were calculated by multiplicatively combining the neural-network color-set scores with the color-saliency scores, and the highest-scoring color sets with six, eight, and ten colors by this metric were selected as the final result. The ten highest-scored and ten lowest-scored color sets with six, eight, and ten colors by this metric are shown in \figurename~\ref{fig:bestworstsets}.

\begin{figure}
\centering
\begin{tabular*}{\columnwidth}{@{\extracolsep{\stretch{1}}}rccc@{}}
\toprule
Rank & Six Colors & Eight Colors & Ten Colors \\
\midrule
1 & \textcolor[HTML]{5790fc}{$\blacksquare$}\textcolor[HTML]{9c9ca1}{$\blacksquare$}\textcolor[HTML]{7a21dd}{$\blacksquare$}\textcolor[HTML]{964a8b}{$\blacksquare$}\textcolor[HTML]{e42536}{$\blacksquare$}\textcolor[HTML]{f89c20}{$\blacksquare$} & \textcolor[HTML]{86c8dd}{$\blacksquare$}\textcolor[HTML]{578dff}{$\blacksquare$}\textcolor[HTML]{1845fb}{$\blacksquare$}\textcolor[HTML]{656364}{$\blacksquare$}\textcolor[HTML]{c849a9}{$\blacksquare$}\textcolor[HTML]{c91f16}{$\blacksquare$}\textcolor[HTML]{ff5e02}{$\blacksquare$}\textcolor[HTML]{adad7d}{$\blacksquare$} & \textcolor[HTML]{94a4a2}{$\blacksquare$}\textcolor[HTML]{92dadd}{$\blacksquare$}\textcolor[HTML]{3f90da}{$\blacksquare$}\textcolor[HTML]{717581}{$\blacksquare$}\textcolor[HTML]{832db6}{$\blacksquare$}\textcolor[HTML]{bd1f01}{$\blacksquare$}\textcolor[HTML]{a96b59}{$\blacksquare$}\textcolor[HTML]{e76300}{$\blacksquare$}\textcolor[HTML]{ffa90e}{$\blacksquare$}\textcolor[HTML]{b9ac70}{$\blacksquare$} \\
2 & \textcolor[HTML]{46a7f9}{$\blacksquare$}\textcolor[HTML]{4d3cf4}{$\blacksquare$}\textcolor[HTML]{908c92}{$\blacksquare$}\textcolor[HTML]{8d4d8e}{$\blacksquare$}\textcolor[HTML]{d82946}{$\blacksquare$}\textcolor[HTML]{ff8f24}{$\blacksquare$} & \textcolor[HTML]{0f9ef7}{$\blacksquare$}\textcolor[HTML]{a0b2ca}{$\blacksquare$}\textcolor[HTML]{3765c9}{$\blacksquare$}\textcolor[HTML]{837f83}{$\blacksquare$}\textcolor[HTML]{724a6f}{$\blacksquare$}\textcolor[HTML]{c33732}{$\blacksquare$}\textcolor[HTML]{fb9f90}{$\blacksquare$}\textcolor[HTML]{ff6709}{$\blacksquare$} & \textcolor[HTML]{add2e8}{$\blacksquare$}\textcolor[HTML]{0bb3fb}{$\blacksquare$}\textcolor[HTML]{6161e9}{$\blacksquare$}\textcolor[HTML]{a56fa5}{$\blacksquare$}\textcolor[HTML]{96208e}{$\blacksquare$}\textcolor[HTML]{e01b21}{$\blacksquare$}\textcolor[HTML]{fd8c15}{$\blacksquare$}\textcolor[HTML]{d5bb8b}{$\blacksquare$}\textcolor[HTML]{8e9381}{$\blacksquare$}\textcolor[HTML]{666765}{$\blacksquare$} \\
3 & \textcolor[HTML]{62afdb}{$\blacksquare$}\textcolor[HTML]{3867cf}{$\blacksquare$}\textcolor[HTML]{8d3663}{$\blacksquare$}\textcolor[HTML]{d61d06}{$\blacksquare$}\textcolor[HTML]{fb970b}{$\blacksquare$}\textcolor[HTML]{8a876f}{$\blacksquare$} & \textcolor[HTML]{60c4fc}{$\blacksquare$}\textcolor[HTML]{4483e5}{$\blacksquare$}\textcolor[HTML]{6926eb}{$\blacksquare$}\textcolor[HTML]{924c84}{$\blacksquare$}\textcolor[HTML]{ce2804}{$\blacksquare$}\textcolor[HTML]{fd7d00}{$\blacksquare$}\textcolor[HTML]{a78a6a}{$\blacksquare$}\textcolor[HTML]{c1c3b9}{$\blacksquare$} & \textcolor[HTML]{9fa7ad}{$\blacksquare$}\textcolor[HTML]{94c4f4}{$\blacksquare$}\textcolor[HTML]{4c8af2}{$\blacksquare$}\textcolor[HTML]{6e32f7}{$\blacksquare$}\textcolor[HTML]{9d6c9f}{$\blacksquare$}\textcolor[HTML]{7d426c}{$\blacksquare$}\textcolor[HTML]{bf241a}{$\blacksquare$}\textcolor[HTML]{df6a09}{$\blacksquare$}\textcolor[HTML]{edc22d}{$\blacksquare$}\textcolor[HTML]{587d63}{$\blacksquare$} \\
4 & \textcolor[HTML]{3397f8}{$\blacksquare$}\textcolor[HTML]{3f49e1}{$\blacksquare$}\textcolor[HTML]{b4acb3}{$\blacksquare$}\textcolor[HTML]{aa5f96}{$\blacksquare$}\textcolor[HTML]{c62832}{$\blacksquare$}\textcolor[HTML]{f1781b}{$\blacksquare$} & \textcolor[HTML]{788387}{$\blacksquare$}\textcolor[HTML]{73ace7}{$\blacksquare$}\textcolor[HTML]{3366fa}{$\blacksquare$}\textcolor[HTML]{a51e9d}{$\blacksquare$}\textcolor[HTML]{a71303}{$\blacksquare$}\textcolor[HTML]{e66738}{$\blacksquare$}\textcolor[HTML]{f9b202}{$\blacksquare$}\textcolor[HTML]{bbb598}{$\blacksquare$} & \textcolor[HTML]{677472}{$\blacksquare$}\textcolor[HTML]{76cdf9}{$\blacksquare$}\textcolor[HTML]{327afe}{$\blacksquare$}\textcolor[HTML]{6025f8}{$\blacksquare$}\textcolor[HTML]{7f499f}{$\blacksquare$}\textcolor[HTML]{ac7cb4}{$\blacksquare$}\textcolor[HTML]{ee2a30}{$\blacksquare$}\textcolor[HTML]{ef9603}{$\blacksquare$}\textcolor[HTML]{cdcf84}{$\blacksquare$}\textcolor[HTML]{93a899}{$\blacksquare$} \\
5 & \textcolor[HTML]{579be7}{$\blacksquare$}\textcolor[HTML]{3341f8}{$\blacksquare$}\textcolor[HTML]{78777c}{$\blacksquare$}\textcolor[HTML]{bd2238}{$\blacksquare$}\textcolor[HTML]{f8a228}{$\blacksquare$}\textcolor[HTML]{b0aa93}{$\blacksquare$} & \textcolor[HTML]{9ebccf}{$\blacksquare$}\textcolor[HTML]{396a8d}{$\blacksquare$}\textcolor[HTML]{5499fc}{$\blacksquare$}\textcolor[HTML]{701ee4}{$\blacksquare$}\textcolor[HTML]{cd022e}{$\blacksquare$}\textcolor[HTML]{fa720c}{$\blacksquare$}\textcolor[HTML]{e6b987}{$\blacksquare$}\textcolor[HTML]{7b7976}{$\blacksquare$} & \textcolor[HTML]{accbe0}{$\blacksquare$}\textcolor[HTML]{4cabf6}{$\blacksquare$}\textcolor[HTML]{195be5}{$\blacksquare$}\textcolor[HTML]{823886}{$\blacksquare$}\textcolor[HTML]{a576a1}{$\blacksquare$}\textcolor[HTML]{e81600}{$\blacksquare$}\textcolor[HTML]{fb9e0c}{$\blacksquare$}\textcolor[HTML]{aeaaa8}{$\blacksquare$}\textcolor[HTML]{71746a}{$\blacksquare$}\textcolor[HTML]{78a05f}{$\blacksquare$} \\
6 & \textcolor[HTML]{519bd4}{$\blacksquare$}\textcolor[HTML]{4358c9}{$\blacksquare$}\textcolor[HTML]{726a7e}{$\blacksquare$}\textcolor[HTML]{ea2425}{$\blacksquare$}\textcolor[HTML]{ff9d47}{$\blacksquare$}\textcolor[HTML]{a9a195}{$\blacksquare$} & \textcolor[HTML]{b5c8c6}{$\blacksquare$}\textcolor[HTML]{3fc9fc}{$\blacksquare$}\textcolor[HTML]{4b8ed4}{$\blacksquare$}\textcolor[HTML]{0d4cfd}{$\blacksquare$}\textcolor[HTML]{b0286e}{$\blacksquare$}\textcolor[HTML]{dd0002}{$\blacksquare$}\textcolor[HTML]{f58201}{$\blacksquare$}\textcolor[HTML]{b19572}{$\blacksquare$} & \textcolor[HTML]{91b1b8}{$\blacksquare$}\textcolor[HTML]{33d3fe}{$\blacksquare$}\textcolor[HTML]{5591f1}{$\blacksquare$}\textcolor[HTML]{115ef9}{$\blacksquare$}\textcolor[HTML]{757b8d}{$\blacksquare$}\textcolor[HTML]{714a8f}{$\blacksquare$}\textcolor[HTML]{cb2b13}{$\blacksquare$}\textcolor[HTML]{fc650e}{$\blacksquare$}\textcolor[HTML]{9c7c64}{$\blacksquare$}\textcolor[HTML]{dbc871}{$\blacksquare$} \\
7 & \textcolor[HTML]{6e7980}{$\blacksquare$}\textcolor[HTML]{53a5eb}{$\blacksquare$}\textcolor[HTML]{3551f6}{$\blacksquare$}\textcolor[HTML]{eb383c}{$\blacksquare$}\textcolor[HTML]{f3af1d}{$\blacksquare$}\textcolor[HTML]{b2b0a4}{$\blacksquare$} & \textcolor[HTML]{8a9fa7}{$\blacksquare$}\textcolor[HTML]{36baf0}{$\blacksquare$}\textcolor[HTML]{0f61ff}{$\blacksquare$}\textcolor[HTML]{8a329c}{$\blacksquare$}\textcolor[HTML]{c9606e}{$\blacksquare$}\textcolor[HTML]{da1701}{$\blacksquare$}\textcolor[HTML]{f4a103}{$\blacksquare$}\textcolor[HTML]{c4c791}{$\blacksquare$} & \textcolor[HTML]{79bbe9}{$\blacksquare$}\textcolor[HTML]{7c80e3}{$\blacksquare$}\textcolor[HTML]{642fe3}{$\blacksquare$}\textcolor[HTML]{a690a4}{$\blacksquare$}\textcolor[HTML]{923f88}{$\blacksquare$}\textcolor[HTML]{da2229}{$\blacksquare$}\textcolor[HTML]{f67b03}{$\blacksquare$}\textcolor[HTML]{ffb575}{$\blacksquare$}\textcolor[HTML]{706d68}{$\blacksquare$}\textcolor[HTML]{b6c1b7}{$\blacksquare$} \\
8 & \textcolor[HTML]{919e9f}{$\blacksquare$}\textcolor[HTML]{695efe}{$\blacksquare$}\textcolor[HTML]{85398c}{$\blacksquare$}\textcolor[HTML]{f51702}{$\blacksquare$}\textcolor[HTML]{fba459}{$\blacksquare$}\textcolor[HTML]{5e6c64}{$\blacksquare$} & \textcolor[HTML]{84adf1}{$\blacksquare$}\textcolor[HTML]{5a67f7}{$\blacksquare$}\textcolor[HTML]{875984}{$\blacksquare$}\textcolor[HTML]{aa1400}{$\blacksquare$}\textcolor[HTML]{e94c14}{$\blacksquare$}\textcolor[HTML]{fab60d}{$\blacksquare$}\textcolor[HTML]{b7baa4}{$\blacksquare$}\textcolor[HTML]{75856e}{$\blacksquare$} & \textcolor[HTML]{adbfb9}{$\blacksquare$}\textcolor[HTML]{7bb5ec}{$\blacksquare$}\textcolor[HTML]{3f7cff}{$\blacksquare$}\textcolor[HTML]{6f08f5}{$\blacksquare$}\textcolor[HTML]{c127ae}{$\blacksquare$}\textcolor[HTML]{b61001}{$\blacksquare$}\textcolor[HTML]{f3501d}{$\blacksquare$}\textcolor[HTML]{f4b843}{$\blacksquare$}\textcolor[HTML]{ae9974}{$\blacksquare$}\textcolor[HTML]{647065}{$\blacksquare$} \\
9 & \textcolor[HTML]{2475f9}{$\blacksquare$}\textcolor[HTML]{635678}{$\blacksquare$}\textcolor[HTML]{eba0e5}{$\blacksquare$}\textcolor[HTML]{c40a09}{$\blacksquare$}\textcolor[HTML]{f07a08}{$\blacksquare$}\textcolor[HTML]{819188}{$\blacksquare$} & \textcolor[HTML]{83a7ba}{$\blacksquare$}\textcolor[HTML]{4e3dea}{$\blacksquare$}\textcolor[HTML]{ab6be7}{$\blacksquare$}\textcolor[HTML]{894890}{$\blacksquare$}\textcolor[HTML]{e22129}{$\blacksquare$}\textcolor[HTML]{f0970b}{$\blacksquare$}\textcolor[HTML]{c1bbaa}{$\blacksquare$}\textcolor[HTML]{70716f}{$\blacksquare$} & \textcolor[HTML]{838e8c}{$\blacksquare$}\textcolor[HTML]{67b5fd}{$\blacksquare$}\textcolor[HTML]{547cf5}{$\blacksquare$}\textcolor[HTML]{bfc0d1}{$\blacksquare$}\textcolor[HTML]{4f39ec}{$\blacksquare$}\textcolor[HTML]{b42b93}{$\blacksquare$}\textcolor[HTML]{dd0e08}{$\blacksquare$}\textcolor[HTML]{6d605a}{$\blacksquare$}\textcolor[HTML]{f89d01}{$\blacksquare$}\textcolor[HTML]{b09d63}{$\blacksquare$} \\
10 & \textcolor[HTML]{55d5e6}{$\blacksquare$}\textcolor[HTML]{53616a}{$\blacksquare$}\textcolor[HTML]{4693c7}{$\blacksquare$}\textcolor[HTML]{c91b1f}{$\blacksquare$}\textcolor[HTML]{ff8c1b}{$\blacksquare$}\textcolor[HTML]{9a9c86}{$\blacksquare$} & \textcolor[HTML]{68c2e2}{$\blacksquare$}\textcolor[HTML]{2a84ff}{$\blacksquare$}\textcolor[HTML]{8f4199}{$\blacksquare$}\textcolor[HTML]{a21d1f}{$\blacksquare$}\textcolor[HTML]{fc3729}{$\blacksquare$}\textcolor[HTML]{f5b75d}{$\blacksquare$}\textcolor[HTML]{74726d}{$\blacksquare$}\textcolor[HTML]{a6a69a}{$\blacksquare$} & \textcolor[HTML]{81abb1}{$\blacksquare$}\textcolor[HTML]{83d8f5}{$\blacksquare$}\textcolor[HTML]{02a3e3}{$\blacksquare$}\textcolor[HTML]{4961ee}{$\blacksquare$}\textcolor[HTML]{734787}{$\blacksquare$}\textcolor[HTML]{c05969}{$\blacksquare$}\textcolor[HTML]{b41e0f}{$\blacksquare$}\textcolor[HTML]{f04005}{$\blacksquare$}\textcolor[HTML]{fda23e}{$\blacksquare$}\textcolor[HTML]{acb089}{$\blacksquare$} \\
\midrule
9991 & \textcolor[HTML]{78d2c1}{$\blacksquare$}\textcolor[HTML]{d78eff}{$\blacksquare$}\textcolor[HTML]{c504ff}{$\blacksquare$}\textcolor[HTML]{a91f69}{$\blacksquare$}\textcolor[HTML]{73762f}{$\blacksquare$}\textcolor[HTML]{90ae20}{$\blacksquare$} & \textcolor[HTML]{51587f}{$\blacksquare$}\textcolor[HTML]{c3a4fc}{$\blacksquare$}\textcolor[HTML]{a039f6}{$\blacksquare$}\textcolor[HTML]{c07c96}{$\blacksquare$}\textcolor[HTML]{825559}{$\blacksquare$}\textcolor[HTML]{748a3e}{$\blacksquare$}\textcolor[HTML]{7dbd0f}{$\blacksquare$}\textcolor[HTML]{47e3aa}{$\blacksquare$} & \textcolor[HTML]{99c5bd}{$\blacksquare$}\textcolor[HTML]{515a77}{$\blacksquare$}\textcolor[HTML]{8f8798}{$\blacksquare$}\textcolor[HTML]{a436e8}{$\blacksquare$}\textcolor[HTML]{ef5ef8}{$\blacksquare$}\textcolor[HTML]{984449}{$\blacksquare$}\textcolor[HTML]{b5b316}{$\blacksquare$}\textcolor[HTML]{5a8f1b}{$\blacksquare$}\textcolor[HTML]{00ef95}{$\blacksquare$}\textcolor[HTML]{00b275}{$\blacksquare$} \\
9992 & \textcolor[HTML]{7d3df3}{$\blacksquare$}\textcolor[HTML]{ce92f4}{$\blacksquare$}\textcolor[HTML]{bd6097}{$\blacksquare$}\textcolor[HTML]{5e6215}{$\blacksquare$}\textcolor[HTML]{8a9e4d}{$\blacksquare$}\textcolor[HTML]{9bccb2}{$\blacksquare$} & \textcolor[HTML]{b6b0f4}{$\blacksquare$}\textcolor[HTML]{8b0df4}{$\blacksquare$}\textcolor[HTML]{c540a7}{$\blacksquare$}\textcolor[HTML]{7b494a}{$\blacksquare$}\textcolor[HTML]{a3a50e}{$\blacksquare$}\textcolor[HTML]{6d7700}{$\blacksquare$}\textcolor[HTML]{649f75}{$\blacksquare$}\textcolor[HTML]{69deaf}{$\blacksquare$} & \textcolor[HTML]{b0bbd6}{$\blacksquare$}\textcolor[HTML]{85879c}{$\blacksquare$}\textcolor[HTML]{5a5a6b}{$\blacksquare$}\textcolor[HTML]{8d56c7}{$\blacksquare$}\textcolor[HTML]{d675f8}{$\blacksquare$}\textcolor[HTML]{9c7811}{$\blacksquare$}\textcolor[HTML]{b0b23a}{$\blacksquare$}\textcolor[HTML]{60e99d}{$\blacksquare$}\textcolor[HTML]{2d7950}{$\blacksquare$}\textcolor[HTML]{41a97f}{$\blacksquare$} \\
9993 & \textcolor[HTML]{39746c}{$\blacksquare$}\textcolor[HTML]{e2a0fa}{$\blacksquare$}\textcolor[HTML]{b54bd1}{$\blacksquare$}\textcolor[HTML]{6c591b}{$\blacksquare$}\textcolor[HTML]{959b4c}{$\blacksquare$}\textcolor[HTML]{85baa9}{$\blacksquare$} & \textcolor[HTML]{649796}{$\blacksquare$}\textcolor[HTML]{436363}{$\blacksquare$}\textcolor[HTML]{ba74fd}{$\blacksquare$}\textcolor[HTML]{ac1eae}{$\blacksquare$}\textcolor[HTML]{ed99d9}{$\blacksquare$}\textcolor[HTML]{747b4d}{$\blacksquare$}\textcolor[HTML]{a7b37a}{$\blacksquare$}\textcolor[HTML]{72e40d}{$\blacksquare$} & \textcolor[HTML]{537a9d}{$\blacksquare$}\textcolor[HTML]{c2cdda}{$\blacksquare$}\textcolor[HTML]{ae3efc}{$\blacksquare$}\textcolor[HTML]{a399a6}{$\blacksquare$}\textcolor[HTML]{f665f0}{$\blacksquare$}\textcolor[HTML]{8a4259}{$\blacksquare$}\textcolor[HTML]{b0756d}{$\blacksquare$}\textcolor[HTML]{dab598}{$\blacksquare$}\textcolor[HTML]{6d6c11}{$\blacksquare$}\textcolor[HTML]{72a421}{$\blacksquare$} \\
9994 & \textcolor[HTML]{bcb2d6}{$\blacksquare$}\textcolor[HTML]{8212e7}{$\blacksquare$}\textcolor[HTML]{cd60f8}{$\blacksquare$}\textcolor[HTML]{a64f5e}{$\blacksquare$}\textcolor[HTML]{8b8a0d}{$\blacksquare$}\textcolor[HTML]{31c487}{$\blacksquare$} & \textcolor[HTML]{92b8b1}{$\blacksquare$}\textcolor[HTML]{5e8594}{$\blacksquare$}\textcolor[HTML]{a41eee}{$\blacksquare$}\textcolor[HTML]{e862ec}{$\blacksquare$}\textcolor[HTML]{952d59}{$\blacksquare$}\textcolor[HTML]{6f651e}{$\blacksquare$}\textcolor[HTML]{c8ca7c}{$\blacksquare$}\textcolor[HTML]{8b9c06}{$\blacksquare$} & \textcolor[HTML]{aebadc}{$\blacksquare$}\textcolor[HTML]{9987ca}{$\blacksquare$}\textcolor[HTML]{8a4ec2}{$\blacksquare$}\textcolor[HTML]{a72872}{$\blacksquare$}\textcolor[HTML]{938610}{$\blacksquare$}\textcolor[HTML]{b6b534}{$\blacksquare$}\textcolor[HTML]{57600c}{$\blacksquare$}\textcolor[HTML]{58855e}{$\blacksquare$}\textcolor[HTML]{76e69f}{$\blacksquare$}\textcolor[HTML]{32b79d}{$\blacksquare$} \\
9995 & \textcolor[HTML]{c7b5f0}{$\blacksquare$}\textcolor[HTML]{bf38b7}{$\blacksquare$}\textcolor[HTML]{c29c13}{$\blacksquare$}\textcolor[HTML]{7e6c09}{$\blacksquare$}\textcolor[HTML]{3f6c59}{$\blacksquare$}\textcolor[HTML]{76c2a7}{$\blacksquare$} & \textcolor[HTML]{a90fff}{$\blacksquare$}\textcolor[HTML]{f978f8}{$\blacksquare$}\textcolor[HTML]{c14ba7}{$\blacksquare$}\textcolor[HTML]{69525e}{$\blacksquare$}\textcolor[HTML]{a19f4a}{$\blacksquare$}\textcolor[HTML]{526e33}{$\blacksquare$}\textcolor[HTML]{90d5b8}{$\blacksquare$}\textcolor[HTML]{6c9a89}{$\blacksquare$} & \textcolor[HTML]{a9a6fd}{$\blacksquare$}\textcolor[HTML]{cec5db}{$\blacksquare$}\textcolor[HTML]{9950e1}{$\blacksquare$}\textcolor[HTML]{bf74ad}{$\blacksquare$}\textcolor[HTML]{b10979}{$\blacksquare$}\textcolor[HTML]{7b4e43}{$\blacksquare$}\textcolor[HTML]{a1a235}{$\blacksquare$}\textcolor[HTML]{6b7504}{$\blacksquare$}\textcolor[HTML]{52986c}{$\blacksquare$}\textcolor[HTML]{71d69f}{$\blacksquare$} \\
9996 & \textcolor[HTML]{0c6973}{$\blacksquare$}\textcolor[HTML]{abaeda}{$\blacksquare$}\textcolor[HTML]{c14ad9}{$\blacksquare$}\textcolor[HTML]{b0a411}{$\blacksquare$}\textcolor[HTML]{6c763f}{$\blacksquare$}\textcolor[HTML]{37ddb0}{$\blacksquare$} & \textcolor[HTML]{7e869d}{$\blacksquare$}\textcolor[HTML]{c4bffb}{$\blacksquare$}\textcolor[HTML]{6645ae}{$\blacksquare$}\textcolor[HTML]{e452fb}{$\blacksquare$}\textcolor[HTML]{994660}{$\blacksquare$}\textcolor[HTML]{c9acae}{$\blacksquare$}\textcolor[HTML]{817a0a}{$\blacksquare$}\textcolor[HTML]{5fc23d}{$\blacksquare$} & \textcolor[HTML]{b3bbdd}{$\blacksquare$}\textcolor[HTML]{575293}{$\blacksquare$}\textcolor[HTML]{9151f1}{$\blacksquare$}\textcolor[HTML]{ba8cff}{$\blacksquare$}\textcolor[HTML]{a473ad}{$\blacksquare$}\textcolor[HTML]{8c4b62}{$\blacksquare$}\textcolor[HTML]{cfa65c}{$\blacksquare$}\textcolor[HTML]{6a7426}{$\blacksquare$}\textcolor[HTML]{6f9d8a}{$\blacksquare$}\textcolor[HTML]{36e7b6}{$\blacksquare$} \\
9997 & \textcolor[HTML]{4f8380}{$\blacksquare$}\textcolor[HTML]{7f3ce2}{$\blacksquare$}\textcolor[HTML]{c279f1}{$\blacksquare$}\textcolor[HTML]{839610}{$\blacksquare$}\textcolor[HTML]{556122}{$\blacksquare$}\textcolor[HTML]{93c2a6}{$\blacksquare$} & \textcolor[HTML]{6d99a5}{$\blacksquare$}\textcolor[HTML]{7002ff}{$\blacksquare$}\textcolor[HTML]{a656f2}{$\blacksquare$}\textcolor[HTML]{e990fd}{$\blacksquare$}\textcolor[HTML]{a0616b}{$\blacksquare$}\textcolor[HTML]{6b6d0c}{$\blacksquare$}\textcolor[HTML]{a3ac38}{$\blacksquare$}\textcolor[HTML]{93d6a8}{$\blacksquare$} & \textcolor[HTML]{a2cac7}{$\blacksquare$}\textcolor[HTML]{8478a3}{$\blacksquare$}\textcolor[HTML]{7635b7}{$\blacksquare$}\textcolor[HTML]{d69aee}{$\blacksquare$}\textcolor[HTML]{d44cff}{$\blacksquare$}\textcolor[HTML]{896409}{$\blacksquare$}\textcolor[HTML]{9cac49}{$\blacksquare$}\textcolor[HTML]{05ef9c}{$\blacksquare$}\textcolor[HTML]{549976}{$\blacksquare$}\textcolor[HTML]{416d5c}{$\blacksquare$} \\
9998 & \textcolor[HTML]{9a14f2}{$\blacksquare$}\textcolor[HTML]{e87ef5}{$\blacksquare$}\textcolor[HTML]{c5479d}{$\blacksquare$}\textcolor[HTML]{656111}{$\blacksquare$}\textcolor[HTML]{989d57}{$\blacksquare$}\textcolor[HTML]{9ec8b8}{$\blacksquare$} & \textcolor[HTML]{afb0c2}{$\blacksquare$}\textcolor[HTML]{80799b}{$\blacksquare$}\textcolor[HTML]{f261f9}{$\blacksquare$}\textcolor[HTML]{7b515c}{$\blacksquare$}\textcolor[HTML]{aa4606}{$\blacksquare$}\textcolor[HTML]{ae9639}{$\blacksquare$}\textcolor[HTML]{79d49d}{$\blacksquare$}\textcolor[HTML]{5a9b81}{$\blacksquare$} & \textcolor[HTML]{0d9b9f}{$\blacksquare$}\textcolor[HTML]{4dd4d9}{$\blacksquare$}\textcolor[HTML]{8149e4}{$\blacksquare$}\textcolor[HTML]{e267f0}{$\blacksquare$}\textcolor[HTML]{9d3c68}{$\blacksquare$}\textcolor[HTML]{838200}{$\blacksquare$}\textcolor[HTML]{b3b456}{$\blacksquare$}\textcolor[HTML]{4c6323}{$\blacksquare$}\textcolor[HTML]{7c9e7b}{$\blacksquare$}\textcolor[HTML]{b5d5b4}{$\blacksquare$} \\
9999 & \textcolor[HTML]{59888d}{$\blacksquare$}\textcolor[HTML]{b0acd9}{$\blacksquare$}\textcolor[HTML]{da28fd}{$\blacksquare$}\textcolor[HTML]{ae9922}{$\blacksquare$}\textcolor[HTML]{6e631f}{$\blacksquare$}\textcolor[HTML]{3ede99}{$\blacksquare$} & \textcolor[HTML]{4c8e96}{$\blacksquare$}\textcolor[HTML]{5b559f}{$\blacksquare$}\textcolor[HTML]{c0bddd}{$\blacksquare$}\textcolor[HTML]{be5dfc}{$\blacksquare$}\textcolor[HTML]{964e60}{$\blacksquare$}\textcolor[HTML]{afa805}{$\blacksquare$}\textcolor[HTML]{7b7922}{$\blacksquare$}\textcolor[HTML]{72c798}{$\blacksquare$} & \textcolor[HTML]{5d89a6}{$\blacksquare$}\textcolor[HTML]{6332de}{$\blacksquare$}\textcolor[HTML]{d0b6e2}{$\blacksquare$}\textcolor[HTML]{fa4cff}{$\blacksquare$}\textcolor[HTML]{aa4e72}{$\blacksquare$}\textcolor[HTML]{bc8982}{$\blacksquare$}\textcolor[HTML]{869402}{$\blacksquare$}\textcolor[HTML]{586e20}{$\blacksquare$}\textcolor[HTML]{9fc45a}{$\blacksquare$}\textcolor[HTML]{8ae0bb}{$\blacksquare$} \\
10000 & \textcolor[HTML]{70c1ae}{$\blacksquare$}\textcolor[HTML]{406869}{$\blacksquare$}\textcolor[HTML]{c7b4fd}{$\blacksquare$}\textcolor[HTML]{cd3de0}{$\blacksquare$}\textcolor[HTML]{ae9d4c}{$\blacksquare$}\textcolor[HTML]{796b18}{$\blacksquare$} & \textcolor[HTML]{48a09b}{$\blacksquare$}\textcolor[HTML]{596586}{$\blacksquare$}\textcolor[HTML]{d694f0}{$\blacksquare$}\textcolor[HTML]{be41ee}{$\blacksquare$}\textcolor[HTML]{84414e}{$\blacksquare$}\textcolor[HTML]{83742f}{$\blacksquare$}\textcolor[HTML]{9dac62}{$\blacksquare$}\textcolor[HTML]{9ccfbb}{$\blacksquare$} & \textcolor[HTML]{7da0ae}{$\blacksquare$}\textcolor[HTML]{ccc7e7}{$\blacksquare$}\textcolor[HTML]{6c516b}{$\blacksquare$}\textcolor[HTML]{f35bf8}{$\blacksquare$}\textcolor[HTML]{c109b7}{$\blacksquare$}\textcolor[HTML]{a17679}{$\blacksquare$}\textcolor[HTML]{cda592}{$\blacksquare$}\textcolor[HTML]{879741}{$\blacksquare$}\textcolor[HTML]{4e7024}{$\blacksquare$}\textcolor[HTML]{36e129}{$\blacksquare$} \\
\midrule
$\max(\min\Delta E_\text{cvd}$) & \textcolor[HTML]{5bb3fa}{$\blacksquare$}\textcolor[HTML]{4d5edc}{$\blacksquare$}\textcolor[HTML]{7c797f}{$\blacksquare$}\textcolor[HTML]{9d311b}{$\blacksquare$}\textcolor[HTML]{c18924}{$\blacksquare$}\textcolor[HTML]{b3c2a7}{$\blacksquare$} & \textcolor[HTML]{c8bee6}{$\blacksquare$}\textcolor[HTML]{8704fd}{$\blacksquare$}\textcolor[HTML]{806f80}{$\blacksquare$}\textcolor[HTML]{c956dd}{$\blacksquare$}\textcolor[HTML]{8e3a3b}{$\blacksquare$}\textcolor[HTML]{d94917}{$\blacksquare$}\textcolor[HTML]{d4b415}{$\blacksquare$}\textcolor[HTML]{74b998}{$\blacksquare$} & \textcolor[HTML]{6f18eb}{$\blacksquare$}\textcolor[HTML]{8d6aed}{$\blacksquare$}\textcolor[HTML]{844693}{$\blacksquare$}\textcolor[HTML]{e8a7da}{$\blacksquare$}\textcolor[HTML]{bc72a6}{$\blacksquare$}\textcolor[HTML]{c24a15}{$\blacksquare$}\textcolor[HTML]{f7882a}{$\blacksquare$}\textcolor[HTML]{7fe593}{$\blacksquare$}\textcolor[HTML]{72ad87}{$\blacksquare$}\textcolor[HTML]{3c7b6a}{$\blacksquare$} \\
\bottomrule
\end{tabular*}

\caption{Best and worst color sets. The ten highest-scored and ten lowest-scored color sets with six, eight, and ten colors, per the metric described in the text, are shown, starting with the set with the best score. Also shown is the color set, of the 10k randomly generated, with the maximum minimum-perceptual distance for each set length. Each color set is ordered by hue angle; ordering data were only collected for the ``better'' set in the survey, so the ordering model is less constained for the ``worst'' sets, so it is not used in this comparison.}
\label{fig:bestworstsets}
\Description[Visualization of best and worst color sets]{The color sets described in the caption are shown, each represented by a set of small squares colored with the colors in the set.}
\end{figure}

\subsection{Final Color Sequences}
\label{sec:final-sequences}

All possible orderings of each of the highest-rated sets were then scored with the trained neural-network color-sequence model, resulting in score ranges of [0.29, 0.90], [0.28, 0.96], [0.29, 0.90], for the six-, eight-, and ten-color sets, respectively. This produced color sequences containing six, eight, and ten colors, which are each deemed to be near-optimal in an aesthetic sense by the neural-network models, given the accessibility restrictions used to generate them. However, there are additional aspects that should be considered to improve accessibility in the case when not all of the colors in the sequences are used. When only a subset of the colors are used, these colors should ideally be chosen to have a larger minimum perceptual distance, including for color-vision deficiencies; have a larger minimum lightness distance, for grayscale display or printing; and be darker, to improve contrast with a white background. To this end, a sequence-accessibility metric is constructed as
\begin{equation}
\text{score}_\text{accessibility} = \frac{1}{N-1}\sum_{i=2}^N\min_{j\in[1, i)} \Delta E_\text{cvd}(i, j) \Delta J'(i, j),
\end{equation}
where $N$ is the number of colors in the sequence, $\Delta E_\text{cvd}(i,j)$ is the perceptual distance---as defined by Equation (\ref{eq:cvd})---between the $i$\textsuperscript{th} and $j$\textsuperscript{th} colors of the sequence, and $\Delta J'(i, j)$ is the lightness distance between the $i$\textsuperscript{th} and $j$\textsuperscript{th} colors of the sequence. This metric accounts for the perceptual-distance and lightness-distance considerations, preferring orderings where there are larger distances between colors earlier in the sequence. Before the metric is applied, white is added as the first color in the sequence, which causes orderings with darker colors toward the beginning of the sequence to be preferred.

An additional consideration comes from linguistic effects on color perception. The seminal work of \citet{Berlin1969} identified eleven basic color terms, single-word terms that unambiguously refer to part of the color gamut. Since these terms are unambiguous, the use of color values that are all described by unique basic color terms makes it easier to refer to such values in written or verbal descriptions, when compared to color values that are described by the same basic color term. Additionally, previous research has shown that categorically-distinct colors can be discriminated between more rapidly than colors in the same category, even if the color pairs have the same perceptual distance \citep{Winawer2007,Witzel2015}. However, color pairs in the same category are generally preferred aesthetically \citep{Schloss2010}, but as the primary purpose of data plots is to convey information, the improved accessibility is chosen over the potential for improved aesthetics. To account for these effects, the probabilistic naming model described in Appendix~\appcolornames{} is applied to the colors in the color sets, and each color is assigned its most-probable basic color term. This information is then used to eliminate all orderings where a basic color term is repeated prior to all unique basic color terms in the color set being used.

Finally, the sequence machine-learning scores and sequence accessibility scores need to be combined. Since the first color in the sequence is the most important for aesthetic reasons, only the remaining orderings that start with the same color as that of the highest-ranked sequence per the machine-learning model are considered. After this data cut and the previous data cut with regard to repeated basic color terms, there are 48, 288, and 23\,040 orderings remaining for the highest-scoring six-color, eight-color, and ten-color sets, respectively. The remaining orderings had sequence accessibility score ranges of [382, 490], [579, 736], [240, 358], for the six-, eight-, and ten-color sets, respectively. The machine-learning scores and accessibility scores are then combined multiplicatively, and the highest-scoring color sequences with six, eight, and ten colors by this metric were selected as the final results, which are shown in Table~\ref{tab:colorcycles}.

\begin{table*}
\caption{Final Color Sequences for Six, Eight, and Ten Colors}
\label{tab:colorcycles}
\begin{minipage}{\textwidth}
\vspace*{-1em}
\begin{center}
\begin{tabular*}{\textwidth}{@{\extracolsep{\stretch{1}}}rrrrrrrrrr@{}}
\toprule
\multicolumn{5}{c}{Six Colors} & \multicolumn{5}{c}{Eight Colors} \\ \cmidrule(r){1-5}\cmidrule(rl){6-10}
& \multicolumn{1}{c}{R} & \multicolumn{1}{c}{G} & \multicolumn{1}{c}{B} & $\min\Delta E_\text{cvd}$ & & \multicolumn{1}{c}{R} & \multicolumn{1}{c}{G} & \multicolumn{1}{c}{B} & $\min\Delta E_\text{cvd}$ \\ \midrule
blue \textcolor[HTML]{5790fc}{$\blacksquare$} & 87 & 144 & 252 & 100.0 & blue \textcolor[HTML]{1845fb}{$\blacksquare$} & 24 & 69 & 251 & 100.0 \\
orange \textcolor[HTML]{f89c20}{$\blacksquare$} & 248 & 156 & 32 & 57.1 & orange \textcolor[HTML]{ff5e02}{$\blacksquare$} & 255 & 94 & 2 & 66.9 \\
red \textcolor[HTML]{e42536}{$\blacksquare$} & 228 & 37 & 54 & 21.3 & red \textcolor[HTML]{c91f16}{$\blacksquare$} & 201 & 31 & 22 & 18.2 \\
purple \textcolor[HTML]{964a8b}{$\blacksquare$} & 150 & 74 & 139 & 21.3 & purple \textcolor[HTML]{c849a9}{$\blacksquare$} & 200 & 73 & 169 & 18.1 \\
gray \textcolor[HTML]{9c9ca1}{$\blacksquare$} & 156 & 156 & 161 & 21.3 & gray \textcolor[HTML]{adad7d}{$\blacksquare$} & 173 & 173 & 125 & 18.1 \\
purple \textcolor[HTML]{7a21dd}{$\blacksquare$} & 122 & 33 & 221 & 20.5 & light blue \textcolor[HTML]{86c8dd}{$\blacksquare$} & 134 & 200 & 221 & 18.1 \\
& & & & & blue \textcolor[HTML]{578dff}{$\blacksquare$} & 87 & 141 & 255 & 18.1 \\
& & & & & gray \textcolor[HTML]{656364}{$\blacksquare$} & 101 & 99 & 100 & 18.1 \\ \midrule
\multicolumn{5}{c}{Ten Colors} \\ \cmidrule(r){1-5}
& \multicolumn{1}{c}{R} & \multicolumn{1}{c}{G} & \multicolumn{1}{c}{B} & $\min\Delta E_\text{cvd}$ \\ \midrule
blue \textcolor[HTML]{3f90da}{$\blacksquare$} & 63 & 144 & 218 & 100.0 \\
orange \textcolor[HTML]{ffa90e}{$\blacksquare$} & 255 & 169 & 14 & 56.8 \\
red \textcolor[HTML]{bd1f01}{$\blacksquare$} & 189 & 31 & 1 & 33.4 \\
gray \textcolor[HTML]{94a4a2}{$\blacksquare$} & 148 & 164 & 162 & 22.3 \\
purple \textcolor[HTML]{832db6}{$\blacksquare$} & 131 & 45 & 182 & 18.3 \\
brown \textcolor[HTML]{a96b59}{$\blacksquare$} & 169 & 107 & 89 & 16.4 \\
orange \textcolor[HTML]{e76300}{$\blacksquare$} & 231 & 99 & 0 & 16.3 \\
tan \textcolor[HTML]{b9ac70}{$\blacksquare$} & 185 & 172 & 112 & 16.1 \\
gray \textcolor[HTML]{717581}{$\blacksquare$} & 113 & 117 & 129 & 16.1 \\
light blue \textcolor[HTML]{92dadd}{$\blacksquare$} & 146 & 218 & 221 & 16.1 \\
\bottomrule
\end{tabular*}

\end{center}
\footnotesize
\emph{Note:} The sequences are ordered from top to bottom, and sRGB values $[0, 255]$ are given. The names shown are the most-probable names based on the color-naming model described in Appendix~\appcolornames. The minimum perceptual distances, $\min\Delta E_\text{cvd}$, take into account color-vision-deficiency simulations.
\end{minipage}
\end{table*}

\section{Discussion}
\label{sec:discussion}

For the final results, all three color sequences start with blue, which is not surprising given that many existing color sequences start with blue and since blue is often a preferred color \citep{Palmer2010}. Similarly, it is not surprising that the linear model, which was used as a cross-check, found that higher mean chroma but lower maximum chroma was preferred; existing advice suggests that more-saturated colors are preferable to a point \citep{Palmer2010}, but this should not be overdone to the point that the colors become gaudy \citep{Tufte1990}. Finally, the lack of correlation between the aesthetic-preference and color-saliency scores is consistent with the findings of \citet{Gramazio2017}.

The various aesthetic and accessibility constraints considered in the above analysis are in some cases competing, so a balance had to be found. For example, maximizing lightness difference for grayscale display directly competes with the limiting of the lightness bounds, which was done both to avoid overemphasizing certain plotted features and to maximize contrast with a white background. The restrictions are also use-case dependent; while very-light colors are not acceptable for line plots or scatter plots, they could be fine when used to indicate areas on a map, particularly if boundaries are drawn in another color. As the goal was to make the resulting color sequences as generally-applicable as possible, the more conservative limits were chosen.

However, other considerations do not compete, so they can be more easily combined. The lack of correlation between aesthetic preference and color saliency is an example of this. While higher color saliency is not a factor in aesthetic preference, it does make it easier to name the colors in the sequence, which, in turn, makes it easier to describe them in written or verbal communication. However, since color saliency is language dependent \citep{Kim2019}, the use of this metric biases the results toward the use of color in the English language. While this might make the results less general, the aesthetic-preference score is arguably already biased toward English, since the survey used to collect the data behind it was conducted in English. Furthermore, the lack of correlation found between aesthetic preference and color saliency suggests any effect along these lines is minor and thus not a significant concern.

Although various accessibility aspects are considered here, including for color-vision deficiencies, color---particularly hue---should still not be used as the sole means of communicating information, when it can be avoided. On data plots, line and marker styles should be varied when feasible, to give an additional factor by which to differentiate separate categories. Furthermore, these additional factors should also be used in written and verbal references to these features, such as in figure captions. While color-vision-deficient individuals have particular difficulty with assigning names to colors, a color-only description is equally useless to anyone viewing a data plot in grayscale. Providing a visual example to go with a written reference, similar to what is generally done in plot legends, is also helpful but still does not replace the need to use an additional factor to describe the referenced features.

\section{Comparison with Existing Color Sequences}
\label{sec:comparison}

The $\Delta E_\text{cvd}$ color distance metric defined in Section~\ref{sec:colorsets} and used to ensure the accessibility of the color sequences developed in the present work can also be used to compare the accessibility of the new color sequences with the accessibility of existing color sequences, particularly those that are the default in commonly-used plotting codes. Since many commonly-used color sequences were, unfortunately, not developed with considerations for color-vision deficiencies in mind, their distinguishability for typical trichromats can also be scored using the standard CAM02-UCS color distance, $\Delta E'$.

Table~\ref{tab:comparison} makes such a comparison to a variety of existing color sequences. For sequences that contain more than ten colors, only the first ten colors are considered here. The defaults from Microsoft Excel,\footnote{2019 v2110, \url{https://www.microsoft.com/}} LibreOffice Calc,\footnote{v7.2, \url{https://www.libreoffice.org/}} Google Sheets,\footnote{Accessed 2021-11-21, \url{https://www.google.com/sheets/about/}} MATLAB,\footnote{R2021b, \url{https://www.mathworks.com/products/matlab.html}} Mathematica,\footnote{v12.0, \url{https://www.wolfram.com/mathematica/}} and R\footnote{v4.1.2, \url{https://www.r-project.org/}} are evaluated. Also compared are three sequences that originated from Tableau Software:\footnote{\url{https://www.tableau.com/}} the ``Category 10'' sequence, which was formerly the default in Tableau and is presently the default in a variety of other plotting codes, such as the Matplotlib Python library \citep{Matplotlib}; the ``Tableau Color Blind'' sequence; and the ``Tableau 10'' sequence, which has been the default in Tableau since version 10.\footnote{\url{https://www.tableau.com/about/blog/2016/7/colors-upgrade-tableau-10-56782}} Other sequences from plotting codes are the ``Seaborn Deep'' sequence, which is the default in the Seaborn Python library;\footnote{v0.11.2, \url{https://seaborn.pydata.org/}} the ``Seaborn Colorblind'' sequence, which is also from the Seaborn library; the default sequence from the Plots.jl Julia library;\footnote{v1.24.3, \url{https://docs.juliaplots.org/}} and the default sequence from Plotly Express.\footnote{v5.4.0, \url{https://github.com/plotly/plotly.py}} The remaining sequences are the sequence published by \citet{Okabe2008} and the ``Bright'' sequence published by \citet{Tol2021}, which both aimed to account for color-vision deficiencies; the ColorBrewer ``Set~1'' sequence \citep{Brewer2003};\footnote{\url{https://colorbrewer2.org/}} and the ``batlowS'' sequence.\footnote{\url{https://doi.org/10.5281/zenodo.5501399}}

From the results in Table~\ref{tab:comparison}, it is clear that many widely-used color sequences are not accessible to individuals with color-vision deficiencies. Color sequences developed with accessibility for color-vision deficiencies in mind, e.g., the Okabe and Ito, ``Tableau Color Blind,'' and R sequences,\footnote{\url{https://developer.r-project.org/Blog/public/2019/11/21/a-new-palette-for-r/}} generally performed better, although this is not universally true, since the ``Seaborn Colorblind'' sequence exhibits poor accessibility after only a few colors. While the ``Tol Bright'' sequence---designed with deuteranopia and protanopia in mind---did not perform particularly well on the $\Delta E_\text{cvd}$ metric, it did perform well for the two types of color-vision deficiency it was designed for, which are both far more common than tritanopia; this is unlike the other sequences that performed poorly on the $\Delta E_\text{cvd}$ metric, which also performed poorly for deuteranopia and protanopia. As far as plotting code defaults go, R and Microsoft Excel fared the best, while Matplotlib---as well as the other plotting codes that use the ``Category 10'' sequence by default---and Mathematica fared the worst, with Seaborn and Plots.jl not faring much better.

\begin{table*}
\footnotesize\renewcommand{\arraystretch}{0.82}\centering
\caption{Color Sequences Comparison}
\label{tab:comparison}
\begin{minipage}{\columnwidth}
\begin{center}
\begin{tabular}{@{}crrcrrcrrcrrcrr@{}}
\toprule
\multicolumn{3}{c}{This Work 6} & \multicolumn{3}{c}{This Work 8} & \multicolumn{3}{c}{This Work 10} & \multicolumn{3}{c}{ColorBrewer Set 1} & \multicolumn{3}{c}{batlowS} \\
\multicolumn{3}{c}{$J' \in [41.3, 76.4]$} & \multicolumn{3}{c}{$J' \in [40.2, 78.5]$} & \multicolumn{3}{c}{$J' \in [41.3, 83.7]$} & \multicolumn{3}{c}{$J' \in [49.1, 97.5]$} & \multicolumn{3}{c}{$J' \in [14.4, 89.6]$} \\ \cmidrule(r){1-3}\cmidrule(rl){4-6}\cmidrule(rl){7-9}\cmidrule(rl){10-12}\cmidrule(l){13-15}
& $\min$ & $\min$ & & $\min$ & $\min$ & & $\min$ & $\min$ & & $\min$ & $\min$ & & $\min$ & $\min$ \\
& $\Delta E'$ & $\Delta E_\text{cvd}$ & & $\Delta E'$ & $\Delta E_\text{cvd}$ & & $\Delta E'$ & $\Delta E_\text{cvd}$ & & $\Delta E'$ & $\Delta E_\text{cvd}$ & & $\Delta E'$ & $\Delta E_\text{cvd}$ \\ \midrule
\textcolor[HTML]{5790fc}{$\blacksquare$} & 100.0 & 100.0 & \textcolor[HTML]{1845fb}{$\blacksquare$} & 100.0 & 100.0 & \textcolor[HTML]{3f90da}{$\blacksquare$} & 100.0 & 100.0 & \textcolor[HTML]{e41a1c}{$\blacksquare$} & 100.0 & 100.0 & \textcolor[HTML]{011959}{$\blacksquare$} & 100.0 & 100.0 \\
\textcolor[HTML]{f89c20}{$\blacksquare$} & 64.5 & 57.1 & \textcolor[HTML]{ff5e02}{$\blacksquare$} & 78.0 & 66.9 & \textcolor[HTML]{ffa90e}{$\blacksquare$} & 64.2 & 56.8 & \textcolor[HTML]{377eb8}{$\blacksquare$} & 65.0 & 44.7 & \textcolor[HTML]{faccfa}{$\blacksquare$} & 78.4 & 72.4 \\
\textcolor[HTML]{e42536}{$\blacksquare$} & 34.5 & 21.3 & \textcolor[HTML]{c91f16}{$\blacksquare$} & 20.0 & 18.2 & \textcolor[HTML]{bd1f01}{$\blacksquare$} & 41.4 & 33.4 & \textcolor[HTML]{4daf4a}{$\blacksquare$} & 45.9 & \textcolor[HTML]{990099}{\textbf{11.4}} & \textcolor[HTML]{828231}{$\blacksquare$} & 50.2 & 35.6 \\
\textcolor[HTML]{964a8b}{$\blacksquare$} & 32.1 & 21.3 & \textcolor[HTML]{c849a9}{$\blacksquare$} & 20.0 & 18.1 & \textcolor[HTML]{94a4a2}{$\blacksquare$} & 26.6 & 22.3 & \textcolor[HTML]{984ea3}{$\blacksquare$} & 33.1 & \textcolor[HTML]{990099}{\textbf{4.7}} & \textcolor[HTML]{226061}{$\blacksquare$} & 31.4 & 23.8 \\
\textcolor[HTML]{9c9ca1}{$\blacksquare$} & 28.8 & 21.3 & \textcolor[HTML]{adad7d}{$\blacksquare$} & 20.0 & 18.1 & \textcolor[HTML]{832db6}{$\blacksquare$} & 26.6 & 18.3 & \textcolor[HTML]{ff7f00}{$\blacksquare$} & 25.3 & \textcolor[HTML]{990099}{\textbf{4.7}} & \textcolor[HTML]{f19d6b}{$\blacksquare$} & 29.4 & 16.8 \\
\textcolor[HTML]{7a21dd}{$\blacksquare$} & 23.9 & 20.5 & \textcolor[HTML]{86c8dd}{$\blacksquare$} & 20.0 & 18.1 & \textcolor[HTML]{a96b59}{$\blacksquare$} & 22.8 & 16.4 & \textcolor[HTML]{ffff33}{$\blacksquare$} & 25.3 & \textcolor[HTML]{990099}{\textbf{4.7}} & \textcolor[HTML]{fdb4b4}{$\blacksquare$} & 14.0 & \textcolor[HTML]{990099}{\textbf{9.0}} \\
& & & \textcolor[HTML]{578dff}{$\blacksquare$} & 20.0 & 18.1 & \textcolor[HTML]{e76300}{$\blacksquare$} & 19.7 & 16.3 & \textcolor[HTML]{a65628}{$\blacksquare$} & 20.6 & \textcolor[HTML]{990099}{\textbf{3.5}} & \textcolor[HTML]{114360}{$\blacksquare$} & 14.0 & \textcolor[HTML]{990099}{\textbf{9.0}} \\
& & & \textcolor[HTML]{656364}{$\blacksquare$} & 20.0 & 18.1 & \textcolor[HTML]{b9ac70}{$\blacksquare$} & 19.2 & 16.1 & \textcolor[HTML]{f781bf}{$\blacksquare$} & 20.6 & \textcolor[HTML]{990099}{\textbf{3.5}} & \textcolor[HTML]{4d734d}{$\blacksquare$} & 14.0 & \textcolor[HTML]{990099}{\textbf{9.0}} \\
& & & & & & \textcolor[HTML]{717581}{$\blacksquare$} & 18.7 & 16.1 & \textcolor[HTML]{999999}{$\blacksquare$} & 20.6 & \textcolor[HTML]{990099}{\textbf{3.5}} & \textcolor[HTML]{c09036}{$\blacksquare$} & 14.0 & \textcolor[HTML]{990099}{\textbf{8.4}} \\
& & & & & & \textcolor[HTML]{92dadd}{$\blacksquare$} & 18.7 & 16.1 & & & & \textcolor[HTML]{175262}{$\blacksquare$} & \textcolor[HTML]{990099}{\textbf{7.5}} & \textcolor[HTML]{990099}{\textbf{5.1}} \\
\toprule
\multicolumn{3}{c}{Microsoft Excel} & \multicolumn{3}{c}{LibreOffice Calc} & \multicolumn{3}{c}{Google Sheets} & \multicolumn{3}{c}{Mathematica} & \multicolumn{3}{c}{Plots.jl} \\
\multicolumn{3}{c}{$J' \in [31.1, 82.9]$} & \multicolumn{3}{c}{$J' \in [25.1, 88.4]$} & \multicolumn{3}{c}{$J' \in [57.7, 87.7]$} & \multicolumn{3}{c}{$J' \in [54.4, 84.2]$} & \multicolumn{3}{c}{$J' \in [61.3, 65.7]$} \\ \cmidrule(r){1-3}\cmidrule(rl){4-6}\cmidrule(rl){7-9}\cmidrule(rl){10-12}\cmidrule(l){13-15}
& $\min$ & $\min$ & & $\min$ & $\min$ & & $\min$ & $\min$ & & $\min$ & $\min$ & & $\min$ & $\min$ \\
& $\Delta E'$ & $\Delta E_\text{cvd}$ & & $\Delta E'$ & $\Delta E_\text{cvd}$ & & $\Delta E'$ & $\Delta E_\text{cvd}$ & & $\Delta E'$ & $\Delta E_\text{cvd}$ & & $\Delta E'$ & $\Delta E_\text{cvd}$ \\ \midrule
\textcolor[HTML]{417ebf}{$\blacksquare$} & 100.0 & 100.0 & \textcolor[HTML]{004586}{$\blacksquare$} & 100.0 & 100.0 & \textcolor[HTML]{4285f4}{$\blacksquare$} & 100.0 & 100.0 & \textcolor[HTML]{5e81b5}{$\blacksquare$} & 100.0 & 100.0 & \textcolor[HTML]{009AFA}{$\blacksquare$} & 100.0 & 100.0 \\
\textcolor[HTML]{e38248}{$\blacksquare$} & 54.7 & 44.4 & \textcolor[HTML]{ff420e}{$\blacksquare$} & 73.0 & 52.0 & \textcolor[HTML]{ea4335}{$\blacksquare$} & 64.8 & 50.4 & \textcolor[HTML]{e19c24}{$\blacksquare$} & 54.2 & 46.9 & \textcolor[HTML]{E36F47}{$\blacksquare$} & 60.8 & 48.5 \\
\textcolor[HTML]{a5a6a7}{$\blacksquare$} & 29.1 & 23.6 & \textcolor[HTML]{ffd320}{$\blacksquare$} & 42.7 & 23.8 & \textcolor[HTML]{fbbc04}{$\blacksquare$} & 39.7 & 24.2 & \textcolor[HTML]{8fb032}{$\blacksquare$} & 21.9 & \textcolor[HTML]{990099}{\textbf{2.1}} & \textcolor[HTML]{3EA44E}{$\blacksquare$} & 45.3 & \textcolor[HTML]{990099}{\textbf{7.2}} \\
\textcolor[HTML]{faba45}{$\blacksquare$} & 19.4 & 14.9 & \textcolor[HTML]{579d1c}{$\blacksquare$} & 35.6 & \textcolor[HTML]{990099}{\textbf{8.0}} & \textcolor[HTML]{34a853}{$\blacksquare$} & 38.0 & \textcolor[HTML]{990099}{\textbf{7.8}} & \textcolor[HTML]{eb6235}{$\blacksquare$} & 21.8 & \textcolor[HTML]{990099}{\textbf{2.1}} & \textcolor[HTML]{C371D2}{$\blacksquare$} & 36.7 & \textcolor[HTML]{990099}{\textbf{7.2}} \\
\textcolor[HTML]{5da3d1}{$\blacksquare$} & 14.0 & 12.8 & \textcolor[HTML]{7e0021}{$\blacksquare$} & 35.6 & \textcolor[HTML]{990099}{\textbf{8.0}} & \textcolor[HTML]{ff6d01}{$\blacksquare$} & 15.5 & \textcolor[HTML]{990099}{\textbf{7.8}} & \textcolor[HTML]{8778b3}{$\blacksquare$} & 13.2 & \textcolor[HTML]{990099}{\textbf{1.1}} & \textcolor[HTML]{AC8E18}{$\blacksquare$} & 24.3 & \textcolor[HTML]{990099}{\textbf{5.1}} \\
\textcolor[HTML]{7ca353}{$\blacksquare$} & 14.0 & \textcolor[HTML]{990099}{\textbf{4.1}} & \textcolor[HTML]{83caff}{$\blacksquare$} & 35.6 & \textcolor[HTML]{990099}{\textbf{8.0}} & \textcolor[HTML]{46bdc6}{$\blacksquare$} & 15.5 & \textcolor[HTML]{990099}{\textbf{7.8}} & \textcolor[HTML]{c56e1a}{$\blacksquare$} & 12.7 & \textcolor[HTML]{990099}{\textbf{1.1}} & \textcolor[HTML]{00AAAE}{$\blacksquare$} & 22.7 & \textcolor[HTML]{990099}{\textbf{5.1}} \\
\textcolor[HTML]{224873}{$\blacksquare$} & 14.0 & \textcolor[HTML]{990099}{\textbf{4.1}} & \textcolor[HTML]{314004}{$\blacksquare$} & 35.6 & \textcolor[HTML]{990099}{\textbf{5.1}} & \textcolor[HTML]{7baaf7}{$\blacksquare$} & 14.5 & \textcolor[HTML]{990099}{\textbf{6.6}} & \textcolor[HTML]{5d9ec7}{$\blacksquare$} & \textcolor[HTML]{990099}{\textbf{11.1}} & \textcolor[HTML]{990099}{\textbf{1.1}} & \textcolor[HTML]{ED5E93}{$\blacksquare$} & 20.1 & \textcolor[HTML]{990099}{\textbf{3.8}} \\
\textcolor[HTML]{964b29}{$\blacksquare$} & 14.0 & \textcolor[HTML]{990099}{\textbf{4.1}} & \textcolor[HTML]{aecf00}{$\blacksquare$} & 16.7 & \textcolor[HTML]{990099}{\textbf{4.7}} & \textcolor[HTML]{f07b72}{$\blacksquare$} & 14.5 & \textcolor[HTML]{990099}{\textbf{6.6}} & \textcolor[HTML]{ffbf00}{$\blacksquare$} & \textcolor[HTML]{990099}{\textbf{11.1}} & \textcolor[HTML]{990099}{\textbf{1.1}} & \textcolor[HTML]{C68225}{$\blacksquare$} & \textcolor[HTML]{990099}{\textbf{10.0}} & \textcolor[HTML]{990099}{\textbf{1.1}} \\
\textcolor[HTML]{616263}{$\blacksquare$} & 14.0 & \textcolor[HTML]{990099}{\textbf{4.1}} & \textcolor[HTML]{4b1f6f}{$\blacksquare$} & 16.7 & \textcolor[HTML]{990099}{\textbf{4.7}} & \textcolor[HTML]{fcd04f}{$\blacksquare$} & \textcolor[HTML]{990099}{\textbf{6.8}} & \textcolor[HTML]{990099}{\textbf{6.0}} & \textcolor[HTML]{a5609d}{$\blacksquare$} & \textcolor[HTML]{990099}{\textbf{11.1}} & \textcolor[HTML]{990099}{\textbf{1.1}} & \textcolor[HTML]{00A98D}{$\blacksquare$} & \textcolor[HTML]{990099}{\textbf{10.0}} & \textcolor[HTML]{990099}{\textbf{1.1}} \\
\textcolor[HTML]{95702e}{$\blacksquare$} & 14.0 & \textcolor[HTML]{990099}{\textbf{4.1}} & \textcolor[HTML]{ff950e}{$\blacksquare$} & 16.7 & \textcolor[HTML]{990099}{\textbf{3.3}} & \textcolor[HTML]{71c287}{$\blacksquare$} & \textcolor[HTML]{990099}{\textbf{6.8}} & \textcolor[HTML]{990099}{\textbf{5.4}} & \textcolor[HTML]{929600}{$\blacksquare$} & \textcolor[HTML]{990099}{\textbf{9.0}} & \textcolor[HTML]{990099}{\textbf{1.1}} & \textcolor[HTML]{8E971E}{$\blacksquare$} & \textcolor[HTML]{990099}{\textbf{9.0}} & \textcolor[HTML]{990099}{\textbf{1.1}} \\
\toprule
\multicolumn{3}{c}{Plotly} & \multicolumn{3}{c}{Seaborn Deep} & \multicolumn{3}{c}{Tableau 10} & \multicolumn{3}{c}{Category 10} & \multicolumn{3}{c}{MATLAB} \\
\multicolumn{3}{c}{$J' \in [54.3, 87.6]$} & \multicolumn{3}{c}{$J' \in [49.4, 78.2]$} & \multicolumn{3}{c}{$J' \in [51.2, 84.7]$} & \multicolumn{3}{c}{$J' \in [45.8, 76.8]$} & \multicolumn{3}{c}{$J' \in [38.4, 79.3]$} \\ \cmidrule(r){1-3}\cmidrule(rl){4-6}\cmidrule(rl){7-9}\cmidrule(rl){10-12}\cmidrule(l){13-15}
& $\min$ & $\min$ & & $\min$ & $\min$ & & $\min$ & $\min$ & & $\min$ & $\min$ & & $\min$ & $\min$ \\
& $\Delta E'$ & $\Delta E_\text{cvd}$ & & $\Delta E'$ & $\Delta E_\text{cvd}$ & & $\Delta E'$ & $\Delta E_\text{cvd}$ & & $\Delta E'$ & $\Delta E_\text{cvd}$ & & $\Delta E'$ & $\Delta E_\text{cvd}$ \\ \midrule
\textcolor[HTML]{636EFA}{$\blacksquare$} & 100.0 & 100.0 & \textcolor[HTML]{4C72B0}{$\blacksquare$} & 100.0 & 100.0 & \textcolor[HTML]{4e79a7}{$\blacksquare$} & 100.0 & 100.0 & \textcolor[HTML]{1f77b4}{$\blacksquare$} & 100.0 & 100.0 & \textcolor[HTML]{0072BD}{$\blacksquare$} & 100.0 & 100.0 \\
\textcolor[HTML]{EF553B}{$\blacksquare$} & 61.9 & 52.5 & \textcolor[HTML]{DD8452}{$\blacksquare$} & 51.4 & 42.4 & \textcolor[HTML]{f28e2b}{$\blacksquare$} & 56.5 & 48.5 & \textcolor[HTML]{ff7f0e}{$\blacksquare$} & 65.7 & 54.1 & \textcolor[HTML]{D95319}{$\blacksquare$} & 63.5 & 49.7 \\
\textcolor[HTML]{00CC96}{$\blacksquare$} & 52.3 & 19.3 & \textcolor[HTML]{55A868}{$\blacksquare$} & 37.2 & \textcolor[HTML]{990099}{\textbf{5.3}} & \textcolor[HTML]{e15759}{$\blacksquare$} & 23.4 & 13.7 & \textcolor[HTML]{2ca02c}{$\blacksquare$} & 46.8 & \textcolor[HTML]{990099}{\textbf{3.4}} & \textcolor[HTML]{EDB120}{$\blacksquare$} & 30.7 & 20.7 \\
\textcolor[HTML]{AB63FA}{$\blacksquare$} & 18.3 & \textcolor[HTML]{990099}{\textbf{1.0}} & \textcolor[HTML]{C44E52}{$\blacksquare$} & 19.4 & \textcolor[HTML]{990099}{\textbf{5.3}} & \textcolor[HTML]{76b7b2}{$\blacksquare$} & 23.4 & 13.7 & \textcolor[HTML]{d62728}{$\blacksquare$} & 26.1 & \textcolor[HTML]{990099}{\textbf{3.4}} & \textcolor[HTML]{7E2F8E}{$\blacksquare$} & 30.7 & \textcolor[HTML]{990099}{\textbf{10.0}} \\
\textcolor[HTML]{FFA15A}{$\blacksquare$} & 18.3 & \textcolor[HTML]{990099}{\textbf{1.0}} & \textcolor[HTML]{8172B3}{$\blacksquare$} & 14.1 & \textcolor[HTML]{990099}{\textbf{2.4}} & \textcolor[HTML]{59a14f}{$\blacksquare$} & 23.4 & \textcolor[HTML]{990099}{\textbf{0.8}} & \textcolor[HTML]{9467bd}{$\blacksquare$} & 26.1 & \textcolor[HTML]{990099}{\textbf{2.0}} & \textcolor[HTML]{77AC30}{$\blacksquare$} & 25.6 & \textcolor[HTML]{990099}{\textbf{7.7}} \\
\textcolor[HTML]{19D3F3}{$\blacksquare$} & 18.3 & \textcolor[HTML]{990099}{\textbf{1.0}} & \textcolor[HTML]{937860}{$\blacksquare$} & 14.1 & \textcolor[HTML]{990099}{\textbf{2.4}} & \textcolor[HTML]{edc948}{$\blacksquare$} & 19.3 & \textcolor[HTML]{990099}{\textbf{0.8}} & \textcolor[HTML]{8c564b}{$\blacksquare$} & 23.7 & \textcolor[HTML]{990099}{\textbf{2.0}} & \textcolor[HTML]{4DBEEE}{$\blacksquare$} & 25.6 & \textcolor[HTML]{990099}{\textbf{7.7}} \\
\textcolor[HTML]{FF6692}{$\blacksquare$} & 17.1 & \textcolor[HTML]{990099}{\textbf{1.0}} & \textcolor[HTML]{DA8BC3}{$\blacksquare$} & 14.1 & \textcolor[HTML]{990099}{\textbf{2.4}} & \textcolor[HTML]{b07aa1}{$\blacksquare$} & 19.3 & \textcolor[HTML]{990099}{\textbf{0.8}} & \textcolor[HTML]{e377c2}{$\blacksquare$} & 22.9 & \textcolor[HTML]{990099}{\textbf{2.0}} & \textcolor[HTML]{A2142F}{$\blacksquare$} & 23.3 & \textcolor[HTML]{990099}{\textbf{7.7}} \\
\textcolor[HTML]{B6E880}{$\blacksquare$} & 17.1 & \textcolor[HTML]{990099}{\textbf{1.0}} & \textcolor[HTML]{8C8C8C}{$\blacksquare$} & 13.1 & \textcolor[HTML]{990099}{\textbf{2.4}} & \textcolor[HTML]{ff9da7}{$\blacksquare$} & 19.3 & \textcolor[HTML]{990099}{\textbf{0.8}} & \textcolor[HTML]{7f7f7f}{$\blacksquare$} & 20.2 & \textcolor[HTML]{990099}{\textbf{2.0}} & & & \\
\textcolor[HTML]{FF97FF}{$\blacksquare$} & 17.1 & \textcolor[HTML]{990099}{\textbf{1.0}} & \textcolor[HTML]{CCB974}{$\blacksquare$} & 13.1 & \textcolor[HTML]{990099}{\textbf{2.4}} & \textcolor[HTML]{9c755f}{$\blacksquare$} & 19.3 & \textcolor[HTML]{990099}{\textbf{0.8}} & \textcolor[HTML]{bcbd22}{$\blacksquare$} & 20.2 & \textcolor[HTML]{990099}{\textbf{2.0}} & & & \\
\textcolor[HTML]{FECB52}{$\blacksquare$} & 15.4 & \textcolor[HTML]{990099}{\textbf{1.0}} & \textcolor[HTML]{64B5CD}{$\blacksquare$} & 13.1 & \textcolor[HTML]{990099}{\textbf{2.4}} & \textcolor[HTML]{bab0ac}{$\blacksquare$} & 19.3 & \textcolor[HTML]{990099}{\textbf{0.8}} & \textcolor[HTML]{17becf}{$\blacksquare$} & 20.2 & \textcolor[HTML]{990099}{\textbf{2.0}} & & & \\
\toprule
\multicolumn{3}{c}{R} & \multicolumn{3}{c}{Seaborn Colorblind} & \multicolumn{3}{c}{Tableau Color Blind} & \multicolumn{3}{c}{Okabe and Ito} & \multicolumn{3}{c}{Tol Bright} \\
\multicolumn{3}{c}{$J' \in [0.0, 84.8]$} & \multicolumn{3}{c}{$J' \in [47.1, 89.6]$} & \multicolumn{3}{c}{$J' \in [42.3, 85.1]$} & \multicolumn{3}{c}{$J' \in [0.0, 90.7]$} & \multicolumn{3}{c}{$J' \in [46.4, 78.4]$} \\ \cmidrule(r){1-3}\cmidrule(rl){4-6}\cmidrule(rl){7-9}\cmidrule(rl){10-12}\cmidrule(l){13-15}
& $\min$ & $\min$ & & $\min$ & $\min$ & & $\min$ & $\min$ & & $\min$ & $\min$ & & $\min$ & $\min$ \\
& $\Delta E'$ & $\Delta E_\text{cvd}$ & & $\Delta E'$ & $\Delta E_\text{cvd}$ & & $\Delta E'$ & $\Delta E_\text{cvd}$ & & $\Delta E'$ & $\Delta E_\text{cvd}$ & & $\Delta E'$ & $\Delta E_\text{cvd}$ \\ \midrule
\textcolor[HTML]{000000}{$\blacksquare$} & 100.0 & 100.0 & \textcolor[HTML]{0173B2}{$\blacksquare$} & 100.0 & 100.0 & \textcolor[HTML]{1170aa}{$\blacksquare$} & 100.0 & 100.0 & \textcolor[HTML]{000000}{$\blacksquare$} & 100.0 & 100.0 & \textcolor[HTML]{4477AA}{$\blacksquare$} & 100.0 & 100.0 \\
\textcolor[HTML]{df536b}{$\blacksquare$} & 68.3 & 50.4 & \textcolor[HTML]{DE8F05}{$\blacksquare$} & 62.4 & 55.8 & \textcolor[HTML]{fc7d0b}{$\blacksquare$} & 66.1 & 54.0 & \textcolor[HTML]{E69F00}{$\blacksquare$} & 80.8 & 77.2 & \textcolor[HTML]{EE6677}{$\blacksquare$} & 51.3 & 25.0 \\
\textcolor[HTML]{61d04f}{$\blacksquare$} & 59.1 & 17.7 & \textcolor[HTML]{029E73}{$\blacksquare$} & 34.2 & 12.8 & \textcolor[HTML]{a3acb9}{$\blacksquare$} & 32.2 & 27.9 & \textcolor[HTML]{56B4E9}{$\blacksquare$} & 56.8 & 49.3 & \textcolor[HTML]{228833}{$\blacksquare$} & 41.2 & \textcolor[HTML]{990099}{\textbf{7.4}} \\
\textcolor[HTML]{2297e6}{$\blacksquare$} & 53.0 & 17.7 & \textcolor[HTML]{D55E00}{$\blacksquare$} & 16.3 & \textcolor[HTML]{990099}{\textbf{10.6}} & \textcolor[HTML]{57606c}{$\blacksquare$} & 18.4 & 16.8 & \textcolor[HTML]{009E73}{$\blacksquare$} & 31.5 & 13.8 & \textcolor[HTML]{CCBB44}{$\blacksquare$} & 34.3 & \textcolor[HTML]{990099}{\textbf{7.4}} \\
\textcolor[HTML]{28e2e5}{$\blacksquare$} & 30.0 & 12.7 & \textcolor[HTML]{CC78BC}{$\blacksquare$} & 16.3 & \textcolor[HTML]{990099}{\textbf{9.9}} & \textcolor[HTML]{5fa2ce}{$\blacksquare$} & 16.7 & 12.1 & \textcolor[HTML]{F0E442}{$\blacksquare$} & 20.8 & 13.8 & \textcolor[HTML]{66CCEE}{$\blacksquare$} & 29.9 & \textcolor[HTML]{990099}{\textbf{7.4}} \\
\textcolor[HTML]{cd0bbc}{$\blacksquare$} & 24.8 & \textcolor[HTML]{990099}{\textbf{8.3}} & \textcolor[HTML]{CA9161}{$\blacksquare$} & 13.0 & \textcolor[HTML]{990099}{\textbf{5.8}} & \textcolor[HTML]{c85200}{$\blacksquare$} & 16.7 & 12.1 & \textcolor[HTML]{0072B2}{$\blacksquare$} & 20.8 & 13.1 & \textcolor[HTML]{AA3377}{$\blacksquare$} & 23.7 & \textcolor[HTML]{990099}{\textbf{7.4}} \\
\textcolor[HTML]{f5c710}{$\blacksquare$} & 24.8 & \textcolor[HTML]{990099}{\textbf{8.3}} & \textcolor[HTML]{FBAFE4}{$\blacksquare$} & 13.0 & \textcolor[HTML]{990099}{\textbf{5.8}} & \textcolor[HTML]{7b848f}{$\blacksquare$} & 14.9 & 12.1 & \textcolor[HTML]{D55E00}{$\blacksquare$} & 20.8 & 13.1 & \textcolor[HTML]{BBBBBB}{$\blacksquare$} & 23.4 & \textcolor[HTML]{990099}{\textbf{7.4}} \\
\textcolor[HTML]{9e9e9e}{$\blacksquare$} & 24.8 & \textcolor[HTML]{990099}{\textbf{8.3}} & \textcolor[HTML]{949494}{$\blacksquare$} & 13.0 & \textcolor[HTML]{990099}{\textbf{5.8}} & \textcolor[HTML]{a3cce9}{$\blacksquare$} & 13.2 & \textcolor[HTML]{990099}{\textbf{11.6}} & \textcolor[HTML]{CC79A7}{$\blacksquare$} & 20.8 & \textcolor[HTML]{990099}{\textbf{11.0}} & & & \\
& & & \textcolor[HTML]{ECE133}{$\blacksquare$} & 13.0 & \textcolor[HTML]{990099}{\textbf{5.8}} & \textcolor[HTML]{ffbc79}{$\blacksquare$} & 13.2 & \textcolor[HTML]{990099}{\textbf{11.6}} & & & & & & \\
& & & \textcolor[HTML]{56B4E9}{$\blacksquare$} & 13.0 & \textcolor[HTML]{990099}{\textbf{5.8}} & \textcolor[HTML]{c8d0d9}{$\blacksquare$} & \textcolor[HTML]{990099}{\textbf{11.2}} & \textcolor[HTML]{990099}{\textbf{8.3}} & & & & & & \\
\bottomrule
\end{tabular}

\end{center}
\smallskip
\emph{Note:} Minimum perceptual distance values $\leq$~12 are set in \textcolor[HTML]{990099}{\textbf{bold purple text}} to indicate potential distinguishability issues.
\end{minipage}
\end{table*}

\section{Conclusions}
\label{sec:conclusions}

In this work, color sequences that balance aesthetics and accessibility were developed using a data-driven approach. Using a survey, data on color-sequence aesthetic preference was collected via crowdsourcing, and these data were used to construct machine-learning aesthetic-preference models. Optimal color sequences were then constructed by combining these aesthetic-preference models with accessibility constraints that address color-vision deficiencies, grayscale printing, minimum contrast with the background, and color saliency. As these are competing constraints, there is no correct answer as to how they should be combined. However, this work aimed to find a balance between aesthetics and various accessibility considerations, such that the resulting color sequences can serve as reasonable defaults in plotting codes, which is why there is only one ``final'' result for each sequence length. In such plotting codes, colors are typically used starting with the first color in the sequence, and the same colors are typically used regardless of the number of items being plotted; while the data being plotted do not have ordinality, only a subset of the colors are used if only a few items are plotted, starting with colors are the start of the sequence, which is why the order of the color sequences matters. The comparison to a collection of existing color sequences shows how the newly-developed color sequences have vastly-improved color-vision-deficiency accessibility over many commonly-used color sequences. The survey data collected, the machine-learning aesthetic-preference models, and the code used to conduct the analysis and produce the results described in this paper have been made available \citep{supplement}.

\section*{Acknowledgments}
The author thanks the thousands of individuals who responded to the color-sequence survey, without whom these results would not exist. This work was made possible thanks to the Colorspacious \citep{Colorspacious}, Matplotlib \citep{Matplotlib}, Numba \citep{Numba}, NumPy \citep{NumPy}, Scikit-learn \citep{Scikit-learn}, SciPy \citep{SciPy}, and TensorFlow \citep{TensorFlow} libraries.

\bibliography{paper.bib}

\appendix

\section{Survey Introduction and Directions}
\label{app:directions}

The survey was introduced to the respondant as follows:\\

\begin{quote}
A color cycle is an ordered set of colors used for plotting categorical
data for visualization. Unfortunately, most existing color cycles are
not colorblind-friendly. This is an issue since a significant fraction
of those viewing a particular plot may be colorblind, especially in
fields with diversity shortcomings. The technical aspect of this can be
resolved by enforcing minimum perceptual distances between colors, both
for normal color vision and for various types of simulated
color vision deficiencies.\\

However, this leaves the aesthetic aspect, which is where
this survey comes in. The goal of this survey is to crowd-source the
information needed to generate aesthetically pleasing color cycles that
are also colorblind-friendly. Randomly generated color sets and cycles,
which have a minimum perceptual distance enforced between colors, are
presented for the user to choose the most pleasing one. These data will
then be used to train a model to generate aesthetically pleasing color
cycles. Additionally, once anonymized, the collected data will be
released under a permissive license.\\
\end{quote}

\noindent The survey directions were presented as follows:\\

\begin{quote}
If you are running software to change your display's color temperature, e.g.,
f.lux, Night Shift, or Redshift, please turn it off before continuing.\\

Each iteration of this survey consists of two steps:\\

First, you'll be presented with two color sets, which are ordered by hue.
Choose the colors you prefer (not the order).\vspace{1em}

\definecolor{light}{HTML}{f5f5f5}
\definecolor{warning}{HTML}{ffdd57}
\definecolor{primary}{HTML}{00d1b2}
\definecolor{info}{HTML}{209cee}
\definecolor{dark}{HTML}{363636}
\definecolor{success}{HTML}{23d160}
\definecolor{link}{HTML}{3273dc}
\definecolor{danger}{HTML}{ff3860}

\begin{center}%
\setlength{\tabcolsep}{2pt}%
\footnotesize%
\textsf{%
\begin{tabular}{ccccccccc}
\colorbox{light}{A} & \colorbox{warning}{B} & \colorbox{primary}{\color{white}C} & \colorbox{info}{\color{white}D} & \,or\, & \colorbox{dark}{\color{white}E} & \colorbox{success}{\color{white}F} & \colorbox{link}{\color{white}G} & \colorbox{danger}{\color{white}H}
\end{tabular}
}
\end{center}\vspace{1em}

Then, you'll be presented with four different permutations of the set you
chose. Choose the order you prefer (the colors are the same).\vspace{1em}

\begin{center}%
\setlength{\tabcolsep}{2pt}%
\footnotesize%
\textsf{%
\begin{tabular}{ccccccccccccccccccc}
\colorbox{primary}{\color{white}1} & \colorbox{info}{\color{white}2} & \colorbox{light}{3} & \colorbox{warning}{4} & \,or\, & \colorbox{warning}{4} & \colorbox{info}{\color{white}2} & \colorbox{primary}{\color{white}1} & \colorbox{light}{3} & \,or\, & \colorbox{light}{3} & \colorbox{primary}{\color{white}1} & \colorbox{warning}{4} & \colorbox{info}{\color{white}2} & \,or\, & \colorbox{info}{\color{white}2} & \colorbox{light}{3} & \colorbox{primary}{\color{white}1} & \colorbox{warning}{4}
\end{tabular}
}
\end{center}\vspace{1em}

Remember, the first few colors are the most important, since they will be used
more often, e.g., a plot with only two elements would only use the first two
colors in a given color cycle.\\

This process will repeat ad infinitum (well, until you stop). If you make a
mistake, keep going; your browser's back button will not work.
\end{quote}

\section{Creating a Maximally-distant Sequence}
\label{app:maxdist}

As an alternative to allowing for aesthetic preference, a color sequence can instead be constructed in a manner that maximizes perceptual distance. Two such methods were proposed in \citet{Glasbey2007}, one that used a sequential search algorithm starting with white as the first color and one that used simulated annealing, both operating in the CIELAB color space. CIELAB was an early attempt to create a perceptually-uniform color space \citep{Fairchild2013}, which has been surpassed in accuracy by multiple more-recent color spaces including CAM02-UCS \citep{Luo2013}. However, neither method took into account color-vision deficiencies, although the possibility of extending the methods to do so was discussed.

Here, we extend the sequential-search-algorithm method to use the perceptual-distance metric mentioned in Section~\ref{sec:colorsets}, $\Delta E_\text{cvd}$. Additionally, we optionally allow the lightness range used to be restricted, although white is still included as the first color, regardless of the maximum allowed lightness. Two color sequences were generated using this method, one without lightness restrictions and one with a lightness restriction of $J' \in [40, 90]$, where $J'$ is the CAM02-UCS lightness parameter. These sequences are shown in Table~\ref{tab:maxdistcycles}, along with the minimum perceptual distance between colors after each color is sequentially added to the sequence. These minimum distances informed the minimum distances used in Section~\ref{sec:randomset}. Note that color sequences with higher minimum perceptual distances than those in Table~\ref{tab:maxdistcycles} can be found, since the simulated-annealing method of \citet{Glasbey2007} found a sequence of 11 colors with a minimum perceptual distance $\sim$10\% higher than that of the sequence found by the corresponding sequential search algorithm.

\begin{table}[b]
\caption{Maximally-distant Color Sequences}
\label{tab:maxdistcycles}
\begin{minipage}{\columnwidth}
\vspace*{-1em}
\begin{center}
\begin{tabular*}{\textwidth}{@{\extracolsep{\stretch{1}}}crrrrcrrrr@{}}
\toprule
\multicolumn{5}{c}{$J' \in [0, 100]$} & \multicolumn{5}{c}{$J' \in [40, 90]$} \\ \cmidrule(r){1-5}\cmidrule(l){6-10}
& \multicolumn{1}{c}{R} & \multicolumn{1}{c}{G} & \multicolumn{1}{c}{B} & $\min\Delta E_\text{cvd}$ & & \multicolumn{1}{c}{R} & \multicolumn{1}{c}{G} & \multicolumn{1}{c}{B} & $\min\Delta E_\text{cvd}$ \\ \midrule
\textcolor[HTML]{000000}{$\blacksquare$} & 0 & 0 & 0 & 100.0 & \textcolor[HTML]{0045fe}{$\blacksquare$} & 0 & 69 & 254 & 67.8 \\
\textcolor[HTML]{2965ff}{$\blacksquare$} & 41 & 101 & 255 & 59.4 & \textcolor[HTML]{9c3a00}{$\blacksquare$} & 156 & 58 & 0 & 61.8 \\
\textcolor[HTML]{a36300}{$\blacksquare$} & 163 & 99 & 0 & 54.0 & \textcolor[HTML]{908e9e}{$\blacksquare$} & 144 & 142 & 158 & 37.5 \\
\textcolor[HTML]{484854}{$\blacksquare$} & 72 & 72 & 84 & 33.1 & \textcolor[HTML]{ffa100}{$\blacksquare$} & 255 & 161 & 0 & 36.1 \\
\textcolor[HTML]{01f700}{$\blacksquare$} & 1 & 247 & 0 & 32.8 & \textcolor[HTML]{6c4b7d}{$\blacksquare$} & 108 & 75 & 125 & 24.2 \\
\textcolor[HTML]{9c9bad}{$\blacksquare$} & 156 & 155 & 173 & 32.2 & \textcolor[HTML]{aacdff}{$\blacksquare$} & 170 & 205 & 255 & 23.5 \\
\textcolor[HTML]{0000a5}{$\blacksquare$} & 0 & 0 & 165 & 28.8 & \textcolor[HTML]{5990ff}{$\blacksquare$} & 89 & 144 & 255 & 22.6 \\
\textcolor[HTML]{5e2000}{$\blacksquare$} & 94 & 32 & 0 & 26.7 & \textcolor[HTML]{ff185a}{$\blacksquare$} & 255 & 24 & 90 & 21.3 \\
\textcolor[HTML]{debba4}{$\blacksquare$} & 222 & 187 & 164 & 21.0 & \textcolor[HTML]{d3baaf}{$\blacksquare$} & 211 & 186 & 175 & 20.6 \\
\textcolor[HTML]{557c67}{$\blacksquare$} & 85 & 124 & 103 & 20.6 & \textcolor[HTML]{25ff82}{$\blacksquare$} & 37 & 255 & 130 & 15.4 \\
\bottomrule
\end{tabular*}

\end{center}
\footnotesize
\emph{Note:} The maximally-distant color sequences were constructed using the sequential method of \citet{Glasbey2007} extended to use CAM02-UCS and color-vision-deficiency simulations. The left half is the result when no lightness restriction is included, while the right half is the result when lightness is restricted to $J'\in[40,90]$.
\end{minipage}
\end{table}

\section{Marker-lightness Analysis}
\label{app:lightness}

Experiment two of \citet{Smart2019} examined how different colors and marker sizes affect the ability to discern different marker shapes on a scatter plot. Scatter plots with $L^* = 50$ gray markers as distractors were shown with two colored markers, which were either the same shape or different shapes. The experiment used a binary-forced-choice design where the research subject was asked whether or not the two colored markers were identical, with instructions to complete the task as quickly and accurately as possible. The authors' analysis considered how changing the marker size and color affected accuracy and concluded that there was no significant effect except for decreased accuracy for markers with $L^* > 92$.

\begin{figure}
\centering
\includegraphics{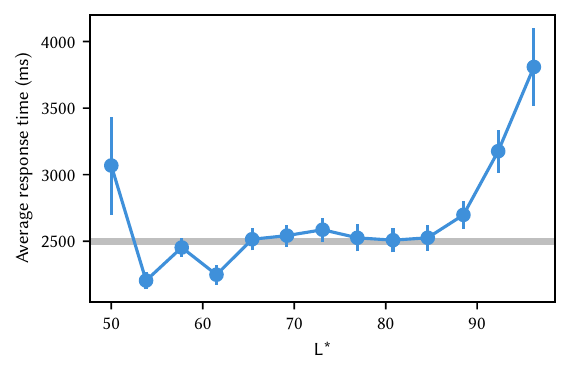}
\caption{Effect of scatter-plot marker lightness on response time. The average response time in milliseconds is shown for markers of various $L^*$, with error bars denoting the standard error. The horizontal gray band shows the overall mean, with a 1-$\sigma$ confidence interval. The data show that the use of colors with $L^*>84.6$ should be avoided to maintain scatter-plot readability.}
\label{fig:marker-lightness}
\Description[Results plot for marker lightness analysis]{A scatter plot with L* lightness on the x-axis and average response time in milliseconds on the y-axis is shown; it includes error bars. Also shown is a narrow horizontal gray band at around 2500 milliseconds. The scatter points at lightness 50, 88.5, 92.3, and 96.2 are significantly above the gray band, and the points at lightness 53.8 and 61.5 are below it. The points at lightness 50 and 96.2 have the largest error bars.}
\end{figure}

The authors of \citet{Smart2019} kindly published their raw data, allowing for reanalysis. Here, the response time is considered instead of the accuracy. Even if two markers can be accurately told apart, the response time should increase when doing so is difficult. Only responses with the smallest marker size used in the experiments, 0.25\textdegree{}, are considered to look at the worst-case scenario; only correct responses are used. Furthermore, responses with response times less than half a second or more than ten seconds are excluded. The results of this analysis are shown in \figurename~\ref{fig:marker-lightness}. Response time significantly increased above the 2.5\,s mean response time for $L^* > 84.6$, suggesting that only colors with lightness less than this threshold should be used in scatter plots to maintain readability. It also decreased somewhat for some of the darker markers, except for $L^*=50$, which had increased response time. The fact that the distractors in the experiment were also of $L^*=50$ is a reasonable explanation for the increased response time for the $L^*=50$ markers.

\section{Color-naming Model}
\label{app:color-names}

Using several million crowdsourced color value--name pairs from the xkcd color survey \citep{Munroe2010}, the authors of \citet{Heer2012} developed a probabilistic naming model for English, which can be used to assign names to color values and to produce color-saliency scores. The saliency scores are meant to quantify the degree to which a color value can be uniquely named, which is a proxy for how easily a color value can be named. However, the construction of the model did not account for color synonyms, and the model uses far more color names than are commonly used. Since the existence of synonymous or rarely-used color names does not negatively affect the nameability of a color value to nearly the same degree that being partway between two commonly-used color names does, the model's lack of synonym merging and the inclusion of rarely-used color names degrades the usefulness of the model's saliency metric for determining how easily a color value can be named. Additionally, the lack of synonym merging can lead to sub-optimal color names for color values near the transition of two color names when multiple synonymous color names are used; for example, the model will use the term ``blue'' for lighter shades of blue than most people would, since while the combination of ``light blue'' and ``sky blue'' contains the majority of the probability for these color values, ``blue'' has a higher probability than either of the other two terms has individually.

The first step in improving the model is to identify and merge synonymous color names. Identifying the most-commonly-used color names as a function of color-gamut volume serves as a starting point for this process. The model discretizes the CIELAB color gamut into 8325 voxels and constructs a color-term count matrix, $T$, with a row for each voxel and a column for each color name. Thus, the set of most-commonly-used color names is constructed by calculating the most-probable color name for each voxel and using the color names that are the most-probable name for at least one voxel, resulting in a set of 33 names. Then, the distances between the probability distributions of each of these color names and the probability distributions of each of the other 152 color names in the model are calculated using the Hellinger distance \citep{LeCam2000}, defined as
\begin{equation}
D_h(w,v) = \sqrt{1-\sum_c\sqrt{p(w|c)p(v|c)}},
\end{equation}
with a summation over all color values $c$ in the discretized CIELAB color gamut and where $w$ and $v$ are color names and $p(w|c)$ is the probability of color name $w$ for color value $c$. Two color names are considered synonyms if the Hellinger distance between their probability distributions is less than 0.25, and the color names are merged by summing together the corresponding columns in the color-term count matrix. When synonyms were merged, the name most commonly provided by the survey respondents was used for the combined data.\footnote{This was not necessarily the same as the name that was the preferred name for the largest number of voxels.}

Next, rarely-used color names were eliminated from the model. To accomplish this, all color names that were not the most-probable name for at least one voxel or not merged with such a name were eliminated by dropping the corresponding columns from the model matrix. The revised model was then used to again find the most-probable color name for each voxel. The numbers of voxels where each of the eleven basic color terms identified by \citet{Berlin1969} was the most-probable color name were tabulated, and ``white'' was found to be the term with the lowest voxel count; as noted in \citet{Heer2012}, this is likely because the xkcd color survey presented the color swatches to be named on a white background. Color names with a voxel count fewer than that of ``white'' were also eliminated by the previously-described method, leaving 28 color names in the final model. This number agrees well with previous estimates of 30 distinct color names used by untrained subjects \citep{Derefeldt1995} and of the sRGB gamut containing 27 categorically-distinct regions of categorically-similar colors \citep{Griffin2019}. The color names in the final model, including synonyms, are shown in Table~\ref{tab:color-names}.

\begin{table}
\centering
\caption{Color-naming-model Names}
\label{tab:color-names}
\begin{minipage}{\columnwidth}
\vspace*{-1em}
\begin{center}
\begin{tabular*}{\columnwidth}{rl}
\toprule
Name & Synonym(s) \\
\midrule
green \\
blue \\
purple & violet \\
red \\
pink \\
yellow \\
orange \\
brown \\
teal & turquoise, blue-green \\
light blue & sky blue \\
gray \\
lime green & lime \\
magenta & fuchsia \\
light green \\
cyan & aqua \\
dark blue & navy blue* \\
dark green & forest green \\
olive \\
lavender & light purple, lilac \\
black \\
tan \\
yellow-green & green-yellow, chartreuse \\
maroon & burgundy \\
salmon \\
peach \\
beige \\
mustard & gold, dark yellow, mustard yellow \\
white \\
\bottomrule
\end{tabular*}
\end{center}
\footnotesize
\emph{Note:} The names in the left column are the 28 names remaining after the model of \citet{Heer2012} was post-processed to merge synonyms and remove rarely-used names. The right column shows names that were determined to be synonymous to the corresponding names in the left column.

\smallskip
* As part of the data pre-processing performed in \citet{Heer2012}, all instances of ``navy'' were converted to ``navy blue,'' so both terms should be considered synonyms.
\end{minipage}
\end{table}

With the revised naming model, the color-saliency metric needs to be renormalized. As is done in \citet{Heer2012}, saliency is defined in terms of negative entropy ($H$),
\begin{equation}
\text{saliency}(c) = -H(p(W|c)) = \sum_{w\in W} p(w|c)\log p(w|c),
\end{equation}
where $p(W|c)$ is the conditional probability of the name of a particular color value $c$, where $p(w|c)$ is the probability of a given color name $w$ for a given color value $c$, and where the summation is over all color names in the model. The saliency is calculated for all voxels in the revised model, giving a saliency range of approximately $-3.158$ to $0$; the minimum and maximum saliency values are then used to normalize the saliency metric to $[0, 1]$.

With the revised model, linear interpolation in the CIELAB color space, instead of binning, is used to calculate color names and saliencies. Colors are named by interpolating the probabilities of each color name separately and using the interpolated probabilities to determine the most-probable name. For saliencies, the voxel saliency values are directly interpolated.

\end{document}